\documentclass[12pt,modern]{aastex63} 

\usepackage{verbatim}
\usepackage{natbib,aas_macros}
\usepackage{hyperref}
\usepackage{graphicx}
\usepackage{xspace}
\usepackage{xcolor}  
\usepackage{rotating} 
\usepackage{breakurl}
\usepackage{amsmath,amssymb,xspace}
\usepackage{fp}
\usepackage{mathtools}

\newcommand{\cchp}{\,CCHP\xspace}                        
\newcommand{\cspi}{\,CSP-I\xspace}                       
\newcommand{\gaia}{\emph{Gaia}\xspace}            
\newcommand{\ho}{$H_{0}$\xspace}                  
\newcommand{\hounits}{\,km\,s$^{-1}$\,Mpc$^{-1}$\xspace}   
\newcommand{\hst}{\emph{HST}\xspace}              
\newcommand{\hstwfciii}{\emph{HST/WFC3}\xspace}    
\newcommand{\hstacs}{\emph{HST/ACS}\xspace}   
\newcommand{\jwst}{\emph{JWST}\xspace}

\newcommand{\sne}{SNe~Ia\xspace}                   
\newcommand{\sn}{SN~Ia\xspace}                   
               
\newcommand{\shoes}{\emph{SHoES}\xspace}            
\newcommand{\ngc}{NGC\,\xspace}
\newcommand{\ic}{IC\,1613\xspace}

\accepted{to the Astrophysical Journal, June 23, 2021}

\newcommand{\hoFb}{69.6}
\newcommand{\hostaterr}{0.8}
\newcommand{\hosyserr}{1.7}
\newcommand{\absmagFb}{-4.054} 
\newcommand{\absmagstatFb}{0.022}
\newcommand{\absmagsysFb}{0.039}

\newcommand{\hoFblong}{$H_0 = \hoFb \pm \hostaterr ~\mathrm{(stat)} \pm\hosyserr ~\mathrm{(sys) ~ }$ \hounits}

\newcommand{\honew}{69.8} 
\newcommand{\honewstaterr}{0.6}
\newcommand{\honewsyserr}{1.6}
\newcommand{\honewlong}{$H_0 = \honew \pm \honewstaterr ~\mathrm{(stat)} \pm\honewsyserr ~\mathrm{(sys)~ }$ \hounits\xspace}
\newcommand{\honewlongnospace}{$H_0 = \honew \pm \honewstaterr ~\mathrm{(stat)} \pm\honewsyserr ~\mathrm{(sys)~ }$ \hounits}

\newcommand{\maserdistmod}{29.397} 
\newcommand{\maserdistmodstat}{0.024} 
\newcommand{\maserdistmodsys}{0.022}
\newcommand{\maserdistmodwerr}{$\mu_0 = \maserdistmod \pm \maserdistmodstat ~\mathrm{(stat)} \pm\maserdistmodsys ~\mathrm{(sys)}~\mathrm{mag}$\xspace}

\newcommand{\Iextinction}{0.025}
\FPeval{\IextinctioROUNDED}{round(\Iextinction,2)}
\FPeval{\Iextinctionerr}{\Iextinction/2}
\FPeval{\IextinctionerrROUNDED}{round(\Iextinction/2,2)}
\FPeval{\IextinctionerrTOTAL}{round((\Iextinctionerr^2+0.01^2)^0.5,2)}

\newcommand{\smcdeb}{18.977}
\newcommand{\smcdebstat}{0.016}
\newcommand{\smcdebsys}{0.028}
\newcommand{\smcdebwerr}{$\mu_0 = \smcdeb \pm \smcdebstat ~\mathrm{(stat)} \pm\smcdebsys ~\mathrm{(sys)}~\mathrm{mag}$\xspace}

\newcommand{\trgbmaglmc}{-4.045}
\newcommand{\trgbmaglmcstat}{0.012} 
\newcommand{\trgbmaglmcsys}{0.034}

\newcommand{\trgbmaggc}{-4.063}
\newcommand{\trgbmaggcstat}{0.07} 
\newcommand{\trgbmaggcsys}{0.11}

\newcommand{\trgbmagmaser}{-4.050}
\newcommand{\trgbmagmaserstat}{0.028} 
\newcommand{\trgbmagmasersys}{0.048}

\newcommand{\trgbmagsmc}{-4.057}
\newcommand{\trgbmagsmcstat}{0.030} 
\newcommand{\trgbmagsmcsys}{0.040}

\newcommand{\trgbmagsculptor}{-4.08}
\newcommand{\trgbmagsculptorstat}{0.06} 
\newcommand{\trgbmagsculptorsys}{0.11}

\newcommand{\trgbmagfornax}{-4.05}
\newcommand{\trgbmagfornaxstat}{0.04} 
\newcommand{\trgbmagfornaxsys}{0.11}

\newcommand{\trgbmaglmcclusters}{-4.085}
\newcommand{\trgbmaglmcclustersstat}{0.05} 
\newcommand{\trgbmaglmcclusterssys}{0.10}

\newcommand{\trgbmagnew}{-4.049}
\newcommand{\trgbmagnewstat}{0.015} 
\newcommand{\trgbmagnewsys}{0.035}

\FPeval\trgbmagnewROUND{round(\trgbmagnew,3)}
\FPeval\trgbmagnewstatROUND{round(\trgbmagnewstat,3)}
\FPeval\trgbmagnewsysROUND{round(\trgbmagnewsys,3)}

 \newcommand{\trgbmagnewlong}{M$_\mathrm{F814W}^\mathrm{TRGB}=\trgbmagnewROUND \pm \trgbmagnewstatROUND ~{\rm (stat)} \pm \trgbmagnewsysROUND  ~{\rm (sys)}~\mathrm{mag}$\xspace}

 \FPeval\trgbmeanmag{ round( (\trgbmaglmc +\trgbmaggc  + \trgbmagmaser + \trgbmagsmc )/4.,2) }
 
 \FPeval\trgbmeaneightmag{ round( (\trgbmaglmc +\trgbmaggc  + \trgbmagmaser + \trgbmagsmc + \trgbmagsculptor + \trgbmagfornax)/7.,2) }
 
\FPeval{\apinffoursevenfive}{round(0.100013,3)} 
\FPeval{\apinffivefivefive}{round(0.0964472,3)} 
\FPeval{\apinfsixzerosix}{round(0.0952613,3)}   
\FPeval{\apinfeightonefour}{round(0.0976345,3)} 
\newcommand{\ZPerr}{0.02}
\newcommand{\EEerr}{0.02}
\newcommand{\Apcorrerr}{0.01}

\newcommand{\trgbobsval}{25.374}
\newcommand{\trgbobsvalstaterr} {0.014} 
\newcommand{\trgbobsvalsyserr} {0.008}
 \FPeval\trgbobsvalROUNDED{round(\trgbobsval,2)} 
 \FPeval\trgbredcorrval{round(\trgbobsval-\Iextinction,2)}
 \FPeval\truetrgbdmod{round(\trgbobsval-\Iextinction-\trgbmaglmc,2)}
 \FPeval\truetrgbdmodMpc{round(10^(\truetrgbdmod/5)/100000,2)}
 \FPeval\trgbobsvalstaterrROUNDED{round(\trgbobsvalstaterr,2)}
 \FPeval\trgbobsvalsyserrROUNDED{round((\trgbobsvalsyserr^2+\ZPerr^2+\EEerr^2+\Apcorrerr^2)^0.5,2)}
\FPeval\trgbcorrsvalsyserrROUNDED{round( (\trgbobsvalsyserr^2+\ZPerr^2+\EEerr^2+\Apcorrerr^2+(\IextinctionerrTOTAL)^2 )^0.5,2)}
 \FPeval\dmodcombinedstaterr{ round( (\trgbobsvalstaterr^2+\trgbmaglmcstat^2)^0.5,2) }
 \FPeval\dmodcombinedsyserr{ round( (\trgbobsvalsyserr^2+\trgbmaglmcsys^2+\ZPerr^2+\EEerr^2+\Apcorrerr^2+\Iextinctionerr^2)^0.5,2) }
 \FPeval\truetrgbdmodMpcupperrdiststat{ 10^( (\truetrgbdmod+\dmodcombinedstaterr) /5)/100000 }
 \FPeval\truetrgbdmodMpclowerdiststat{ 10^( (\truetrgbdmod-\dmodcombinedstaterr) /5)/100000 }
 \FPeval\truetrgbdmodMpcstaterr{ round( 0.5*(\truetrgbdmodMpcupperrdiststat - \truetrgbdmodMpclowerdiststat) ,2) }
 \FPeval\truetrgbdmodMpcupperrdistsys{ 10^( (\truetrgbdmod+\dmodcombinedsyserr) /5)/100000 }
 \FPeval\truetrgbdmodMpclowerdistsys{ 10^( (\truetrgbdmod-\dmodcombinedsyserr) /5)/100000 }
 \FPeval\truetrgbdmodMpcsyserr{ round( 0.5*(\truetrgbdmodMpcupperrdistsys - \truetrgbdmodMpclowerdistsys) ,2) }

\graphicspath{{./}{figures/}}

\shorttitle{TRGB Calibration Update}
\shortauthors{Freedman}

\begin{document}
\title{\textit{Measurements of the Hubble Constant: Tensions in Perspective \protect\footnote{Based on observations made with the NASA/ESA Hubble Space Telescope, obtained at the Space Telescope Science Institute, which is operated by the  Association of Universities for Research in Astronomy, Inc., under NASA contract NAS 5-26555. These observations are associated with programs \#13472 \#13691, \#9477 and \#10399. }}}

\bigskip

\author[0000-0003-3431-9135]{Wendy~L.~Freedman}\affil{Department of Astronomy \& Astrophysics \& Kavli Institute for Cosmological Physics, University of Chicago, 5640 South Ellis Avenue, Chicago, IL 60637, USA}
\email{wfreedman@uchicago.edu}

\begin{abstract}

Measurement of the  distances to nearby galaxies have improved rapidly in recent decades.  The ever-present challenge is to reduce systematic effects, especially as greater distances are probed, and the uncertainties become larger. 
In this paper, we combine several recent calibrations of the Tip of the Red Giant Branch (TRGB) method.  These calibrations are internally self-consistent at the 1\% level. New \gaia Early Data Release 3 (EDR3) data provide an additional consistency check, at a (lower) 5\% level of accuracy,  a result of the  well-documented \gaia angular covariance bias. The updated TRGB calibration applied to a distant sample of Type Ia supernovae from the Carnegie Supernova Project results in a value of the Hubble constant of \honewlongnospace.
No statistically significant difference is found between  the value of \ho  based on the TRGB  and that determined from measurements of the cosmic microwave background. The TRGB results are also consistent to within 2$\sigma$ with the SHoES and {\it Spitzer} plus \hst Key Project Cepheid calibrations.   The TRGB results alone do not demand additional new physics beyond the standard ($\Lambda$CDM) cosmological model. They have the advantage of simplicity of the underlying physics (the core He flash) and small systematic uncertainties (from extinction, metallicity and crowding). Finally, the strengths and weaknesses of both the TRGB and Cepheids are reviewed, and prospects for addressing the current discrepancy with future \gaia, \hst and \jwst  observations are discussed. Resolving this discrepancy is essential for ascertaining if the claimed tension in \ho between the locally-measured and the CMB-inferred value is physically motivated.

\end{abstract}


\section{Introduction}
\label{sec:intro}

Over the last decade, the unprecedented increase in accuracy obtained by a broad range of independent cosmological experiments and observations has provided striking and compelling support for our current standard $\Lambda$ Cold Dark Matter ($\Lambda$CDM) model. This concordance cosmology has been remarkably successful in explaining an even wider range of observations, from the exquisite precision in recent measurements of fluctuations in the temperature and polarization of the cosmic microwave background (CMB) radiation \citep{planck_2018, ACT_2020}  to observations of large-scale structure and matter fluctuations in the universe (e.g.,  baryon acoustic oscillations (BAO), \citeauthor{macaulay_2019}  2019). 

However, as the accuracy of both the observations and the tests of $\Lambda$CDM have improved, a number of discrepancies have been noted. The most apparently significant of these is the claim of a tension between competing values of the Hubble constant (\ho), where the discrepancy is currently estimated to be at the 5 to 6 sigma level \citep{riess_2021, divalentino_2021} between the local values of \ho and those derived from models of the CMB.\footnote{As noted by \citet{feeney_2018}, the true tension between the Planck and the SHoES results depends on accurate knowledge of the tails of the likelihoods of the two distributions, rather than assuming them to be Gaussian. The significance of the current tension also depends on the assumption that all sources of uncertainty have been recognized and accounted for.}  This claimed tension suggests that the universe at present is expanding about 8\% faster than predicted assuming the $\Lambda$CDM model, which, if confirmed, could provide evidence for cracks in the standard model, offering the exciting opportunity for discovering new physics.  Confirming the reality of the \ho tension could have significant consequences for both fundamental physics and modern cosmology.\footnote{For a different perspective on the \ho tension, see the recent review by \citet{linder_2021}.} The implications of an accurate value of \ho are of interest, however, independently of how the tension is ultimately resolved: providing independent confirmation of the standard cosmological model would also be a critical result.

As apparent fissures in the standard model have been emerging, there are also indications that there may be cracks that need attention in the local distance scale as well. For example, the Tip of the Red Giant Branch (TRGB) method and the Cepheid distance scale result in differing values of \ho = 69.6 $\pm$ 1.9 km/sec/Mpc, \citep[][hereafter, F19, F20]{freedman_2019, freedman_2020} for the TRGB and  73.2 $\pm$ 1.3 \citep[][hereafter, R21]{riess_2021} for the Cepheids. This divergence raises the question of whether the purported tension is instead being driven by yet-to-be-revealed systematic errors in the local Cepheid data rather than in the cosmological models. 

A number of measurements of \ho calibrated locally (referred to as late-time estimates) exhibit reasonable agreement to within their quoted uncertainties, generally falling in the range of 70-76 \hounits \citep{freedman_2012, riess_2016, riess_2019, huang_2020,  kourkchi_2020, reid_2019, freedman_2019, freedman_2020, pesce_2020, khetan_2021,blakeslee_2021}.  
In contrast, (early-time) estimates of \ho based on measurements of  fluctuations in the temperature and polarization of the cosmic microwave background (CMB) from Planck and ACT+WMAP  \citep{planck_2018,ACT_2020} consistently yield lower values of \ho = 67.4 $\pm$0.5 and 67.6 $\pm$ 1.1 \hounits, respectively, both adopting the current standard $\Lambda$CDM model. Measurements of fluctuations in the matter density or baryon acoustic oscillations \citep[e.g.,][]{aubourg_2015, macaulay_2019}  also result in similar (low) values, if the absolute scale is set by the sound horizon measurement from the CMB {or by Big Bang nucleosynthesis (BBN) constraints, also based on sound horizon physics}. 

High values of \ho were initially obtained from time-delay measurements of strong  gravitational lensing \citep[][]{Suyu_2017, wong_2020}, with \ho = 73$^{+1.7}_{-1.8}$  \hounits, apparently consistent with the Cepheid measurements. However, recent detailed consideration of the assumptions in the modeling of the lens mass distribution \citep{birrer_2020, birrer_treu_2020} leads to a much lower value of  the Hubble constant, as well as  a significantly larger value of the uncertainty, \ho = 67.4$^{+4.1}_{-3.2}$ \hounits, currently consistent with the CMB and TRGB measurements. 

The debate over the value of the Hubble constant is clearly not yet over. And with the high precision of current CMB measurements, the requirement for greater accuracy in the local value of \ho has grown substantially. 
Given the importance of this question for fundamental physics and for cosmology, and given the history of \ho, and the century-long effort to address a multiplicity of systematic effects, it is essential that rigorous tests be undertaken to investigate the possibility that remaining (potentially unknown) systematic errors are responsible for driving the controversy.

The TRGB method has emerged as one of the most precise and accurate means of measuring distances in the local universe. The TRGB is an excellent standard candle, as an unambiguous signpost of the 
core helium-flash luminosity at the end phase of red giant branch (RGB) evolution for low-mass stars  \citep[e.g.,][]{lee_1993, rizzi_2007, salaris_2002, madore_2009, freedman_2019, jang_2021}. Empirically, observed color-magnitude diagrams of the halos of nearby galaxies reveal a sharp discontinuity at a well-defined luminosity. 

In F19 we presented a determination of \ho based on TRGB distances to 15 galaxies that were hosts to 18 Type Ia supernovae (\sne). $I$-band TRGB distances were measured using \hst Advanced Camera for Surveys ({\it ACS}) data targeting the halo regions of nearby galaxies, and then applied to a sample of 99 significantly more distant \sne (out to z = 0.08) that were observed as part of the Carnegie Supernova Project, and published in \citet{krisciunas_2017}.  This TRGB calibration was  updated slightly in F20, yielding a value of \hoFblong. 
To date, the TRGB is the only method with comparable numbers of galaxies in its calibration relative to Cepheids; the \ho calibration of \citet[][hereafter R16, R19]{riess_2016, riess_2019} is based on the Cepheid distances to 19 galaxies.  
Ten of the galaxies in the F19 and F20  TRGB sample also have independent Cepheid distances, an order of magnitude greater number than for Miras \citep{huang_2020} or the maser technique \citep{pesce_2020}, in both cases for which only a single galaxy is available for comparison with Cepheids.

The immediate goal of this paper is to update the F20 TRGB calibration  of \ho,  which was based solely on a geometric distance to the Large Magellanic Cloud (LMC). In the interim, a number of detailed new studies of the giant branch population in our own and several nearby galaxies can now provide new and independent calibrations of the TRGB. Five independent calibrations are examined in this paper.   These include:
\begin{enumerate}
    \item Observations of the TRGB in the outer halo of the maser galaxy, \ngc 4258 \citep{jang_2021}. 
    \item Observations of TRGB stars in  46 Galactic globular clusters spanning a range of metallicities \citep{cerny_2021a}, calibrated via a Detached Eclipsing Binary (DEB) distance to $\omega$ Cen.
    \item A new  geometric distance to the Small Magellanic Cloud (SMC) based on an augmented sample of 15 DEBs \citep{graczyk_2020}, incorporating the updated reddening and extinction maps of \citet{skowron_2021}, together with an updated measurement of the TRGB magnitude by tt  (2021, in prep). 
    \item  A re-analysis of the OGLE-III data for the LMC by \citet{hoyt_2021b}, incorporating the updated reddening and extinction maps of \citet{skowron_2021}.
    \item New Magellan imaging data for two Milky Way dwarf spheroidal  galaxies, Sculptor
    (Tran et al. 2021) and Fornax (Oakes et al. 2021), as well as \hstacs published data for four LMC globular clusters \citep{olsen_1998} provide an additional check on the calibration of the TRGB zero point. 
\end{enumerate}

A second goal of this paper is to  examine and inter-compare recent calibrations of the TRGB and Cepheid distance scales, and finally, a third goal is to assess the  significance of the tension in \ho, as it currently stands. 

The outline of this paper is as follows.   
In \S \ref{sec:trgbupdate} we describe the recent calibrations of the TRGB; in \S \ref{sec:hotrgb} we discuss the implications of these results in the context of the determination of \ho; in \S \ref{sec:hocepheid} we summarize recent calibrations of the Cepheid Leavitt law. We then compare the \ho values in \S \ref{sec:comparison}, and finally, in \S \ref{sec:discussion} we discuss the current status, strengths and weaknesses in the TRGB and Cepheid distance scales, before comparing our results with other methods in \S \ref{sec:ho_recent} and summarizing our results in \S \ref{sec:summary}.

In brief, based on four independent calibrations of the TRGB absolute magnitude, we find \trgbmagnewlong, leading to a value of \honewlongnospace.  Accurate calibration of the extragalactic distance scale remains a challenging endeavor, and  $<$1\% measurements of the CMB set a high (and currently not attainable) bar for the local distance scale to match. The discrepancy in local (TRGB versus Cepheid) measurements suggests that there are issues in the local distance scale that need to be understood before we can unambiguously make extraordinary claims like new physics.

\section{Absolute Calibration of the TRGB}
\label{sec:trgbupdate}

As can be seen from \autoref{tab:trgbzp} the value of the absolute $I$-band magnitude of the TRGB has remained quite stable over the 30 years in which it has been measured, generally falling within the range of M$_I$ = -4.00 to -4.05 mag [at (V-I)$_o$ = 1.6 mag].

\vspace{2em}
\begin{deluxetable}{l l}[h]
\tablecolumns{2}
\tabletypesize{\small}
\tablecaption{Absolute $I$-band TRGB calibrations \label{tab:trgbzp}}
\tablehead{
\colhead{$M^{I}_{\rm TRGB}$}\tablenotemark{a}& \colhead{Reference}  }
\startdata
-4.0 $\pm 0.1$ & \citet{lee_1993} \\
-4.05  \tablenotemark{b} & \citet{rizzi_2007}  \\
-4.04 \tablenotemark{c} & \citet{bellazzini_2008}   \\
-4.05 $\pm 0.02$ $\pm 0.10$ & \citet{tammann_2008}   \\
-4.01  & \citet{bono_2008a}   \\
-4.03 \tablenotemark{d} & \citet{madore_2009}   \\
-4.02 $\pm 0.06$ \tablenotemark{e} & \citet{jang_lee_2017a}  \\
-4.01 $\pm 0.04$ & \citet{reid_2019}   \\
-3.97 $\pm$ 0.046 & \citet{yuan_2019} \\
-4.05 $\pm 0.02$ $\pm 0.04$ & \citet{freedman_2020}\\
-4.04 $\pm 0.01$ $\pm 0.03$ \tablenotemark{f} & \citet{hoyt_2021b} : LMC \\
-4.05 $\pm$ 0.03 $\pm 0.04$ \tablenotemark{g} & \citet{hoyt_2021b} : SMC \\
\hline \hline
\enddata
\tablenotetext{a}{~At (V-I) = 1.6 mag unless otherwise noted}
\tablenotetext{b}{~-4.05 + 0.217 $\times$ [(V-I) -1.6]}
\tablenotetext{c}{~-3.939 - 0.194 $\times$ (V-I) + 0.08 $\times$ (V-I)$^2$}
\tablenotetext{d}{~-4.05 + 0.2 $\times$ [(V-I) - 1.5]}
\tablenotetext{e}{~-4.016 + 0.091 $\times$[(V-I)$_0$  - 1.5]$^2$ - 0.007 $\times$ [(V-I)$_0$ - 1.5] }
\tablenotetext{f}{~1.60 $< (V - I)_0 <$ 1.95 mag}
\tablenotetext{g}{~1.45 $< (V - I)_0 <$ 1.65 mag}

\end{deluxetable}

In this section, we present  a summary of several independent calibrations of the TRGB that have become available since the \citet{freedman_2020} calibration, which was based solely on the DEB distance to the LMC. Importantly, these calibrations 
are based on very different methods for measuring absolute distances, including a geometric maser technique, geometric parallaxes and geometric DEB distances.  In \S \ref{sec:trgbadopt}, we combine all of these results to obtain an updated calibration of the TRGB. These results are summarized in Table \ref{tab:trgbcal} in \S \ref{sec:trgbadopt}.

\subsection{The Megamaser Galaxy \ngc 4258}
\label{sec:n4258}

The nearby  spiral galaxy   \ngc 4258, at a distance of 7.6 Mpc, is an excellent target for providing a high-accuracy calibration of the TRGB. It  is host to a sample of H$_2$O megamasers, rotating within a highly-inclined 
(87$^\circ$) accretion disk about a supermassive black hole, from which a geometric distance to the galaxy can be measured \citep[see][]{humphreys_2013, reid_2019}. The most recent geometric distance to \ngc 4258 is \maserdistmodwerr \citep{reid_2019}, a 1.5\% measurement. 

The most extensive study  of the TRGB in \ngc 4258 has been published  by \citet{jang_2021}. This measurement is based on a set of 15 archival \hstacs fields covering 54 square arcmin, located near the minor axis in the dust- and gas-free outer halo  of the galaxy. 
The analysis was further confined  primarily to regions at a de-projected semi-major axis distance of $>$14 arcmin ($\sim$ 30 kpc) from the center of the galaxy. 
The RGB stars at this large distance are well-separated from each other, and are  demonstrably free from  crowding/blending effects. Moreover, these halo RGB stars are relatively blue and metal poor, and do not exhibit a wide range in color/metallicity. The wide areal coverage results in a well-populated giant branch with about 3,000 red giant stars one-magnitude below the tip itself.  As described in detail in  \citeauthor{jang_2021}, extensive tests for systematics were undertaken; for example,  using artificial stars; comparing DOLPHOT and DAOPHOT photometry; and comparing results using  different point spread functions, sky-fitting parameters, and radial spatial cuts. Moreover, the \hstacs data used for this study are on the F814W flight magnitude system used in the F19 study, and thus do not require a photometric transformation as for the case of the LMC zero point.   
\citeauthor{jang_2021} obtain a TRGB zero-point of M$_{814}^{TRGB}$ = \trgbmagmaser \ $\pm$ \trgbmagmaserstat \ $\pm$ \trgbmagmasersys, using the maser distance determined by \citet{reid_2019}. 
A detailed description of the error budget and the adopted statistical and systematic uncertainties are given in \S6 and Table 4 of \citeauthor{jang_2021} This independent TRGB calibration agrees to better than 1\%  with the value of M$_{814}$ = -4.054  mag found earlier by F20, as well as that of -4.045 mag measured by \citet{hoyt_2021b}, as described in \S \ref{sec:LMC}.


Alternatively, if we instead  determine the distance to \ngc~4258  based on the  LMC TRGB calibration of  \citet{hoyt_2021b}, given the measured apparent TRGB magnitude of m$_{814,o}^{N4258}= 25.347 \pm 0.014 \pm 0.005$ \citep{jang_2021}, we find a distance modulus of $\mu_o$ =  29.392 $\pm$ 0.018 $\pm$ 0.032 mag.
The agreement with the maser distance  of 29.397 $\pm$ 0.033 mag \citep{reid_2019} is at a level of better than 1\%,  differing by  $<$0.2$\sigma$.  In contrast, we note that a  Cepheid calibration of the distance to \ngc 4258 does not as yield good agreement with that of the maser distance.  As recently described in \citet{efstathiou_2020}, a calibration of the Cepheid distance to \ngc 4258 based on the LMC differs from the maser distance by  2.0-3.5$\sigma$, depending on the adopted correction for metallicity.
The Milky Way and \ngc~4258 metallicities are very similar, however, and should be independent of a metallicity effect. If instead, the Milky Way is adopted as the anchor galaxy to determine the Cepheid distance to \ngc~4258, a distance modulus of 29.242 $\pm$ 0.052 is obtained, which differs from the maser distance by 7\% at a 2$\sigma$ level of significance. We defer a discussion of the implications of these differences to  \S\ref{sec:comparison}. 

Finally, we note that the location of the fields studied by \citet{jang_2021} in the outer halo of \ngc 4258 is optimal for avoiding  dust and gas, as well as being separated from the high surface brightness galactic disk, thereby minimizing the level of systematic effects that  plague efforts to measure the TRGB in the star-forming region of the disk of this galaxy, issues not considered, for example, in \citet{macri_2006, reid_2019}. 

\subsection{Galactic Globular Clusters}
\label{sec:GC}

A second and completely independent method for calibrating the TRGB uses photometry of well-measured giant branches in globular clusters within our own Milky Way. Collectively, the Milky Way globular clusters span a wide range in metallicity, which overlaps well with those measured for giant stars in the halos of nearby, resolved galaxies. 

This approach to calibrating the TRGB was first carried out by \citet{da_costa_armandroff_1990}, using CCD imaging data for six globular clusters. That calibration, for which distances were obtained using theoretical horizontal branch models from \citet{lee_demarque_zinn_1990} (to calibrate the luminosities of RR Lyrae stars),  formed the basis of the \citet{lee_1993} early application of the TRGB method to the extragalactic distance scale. A decade later, \citet{ferraro_1999} assembled a homogenous sample of 60 globular clusters,  adopting the level of the theoretical zero-age horizontal branch  as the basis from which to measure absolute distances. As these authors noted, the advantage of the horizontal branch is the simplicity of the measurement as compared to RR Lyrae stars, for which variability and evolutionary effects need to be accounted for, and for which uncertainties due to metallicity still remain. \citet{bellazzini_2001, bellazzini_2004} based their calibration on observations of the two populous globular clusters, $\omega$ Centauri and 47 Tucanae, calibrated using a DEB distance to $\omega$ Cen \citep{thompson_2001}, and an average of literature distances for 47 Tuc. Subsequently, \citet{rizzi_2007}  based their distances on the well-developed horizontal branches of five Local Group galaxies (\ic, \ngc 185, Fornax, Sculptor and M33) spanning a range in metallicities of -1.74 $<$ [Fe/H] $<$ -1.02 dex. 

In a recent study, \citet{cerny_2021a}  have analyzed a  sample  of  46 low-reddening [E(B-V) $<$ 0.25 mag]
Milky Way globular clusters with uniformly reduced photometry available from \citet{stetson_2019} and through the Canadian Astronomy Data Center (CADC).\footnote{The Stetson catalog is based on a collection of about 90,000 images for 48 clusters, all having $UBVRI$  photometry, for which a comparison of the different data sets constrains the photometric zero-point uncertainties at the millimag level. Eleven of those clusters did not meet the  \citet{cerny_2021a}  low-reddening criterion. \citeauthor{cerny_2021a}  
expanded the Stetson catalog to incorporate nine additional low-reddening clusters with $BVI$ photometry alone, archived at the CADC, and analyzed with the same  DAOPHOT/ALLFRAME software \citep{stetson_1987, stetson_1994}.} This 46-cluster catalog was then cross-matched to the \gaia Data Release 2 (DR2) database, and membership for these clusters was determined using the DR2 proper motion data and a Gaussian-mixture-model clustering algorithm. Preliminary $E(B-V)$ reddening estimates and initial distance estimates were taken from \citet{harris_1996, harris_2010}.

\begin{figure*} 
 \centering
\includegraphics[width=1.0\textwidth]{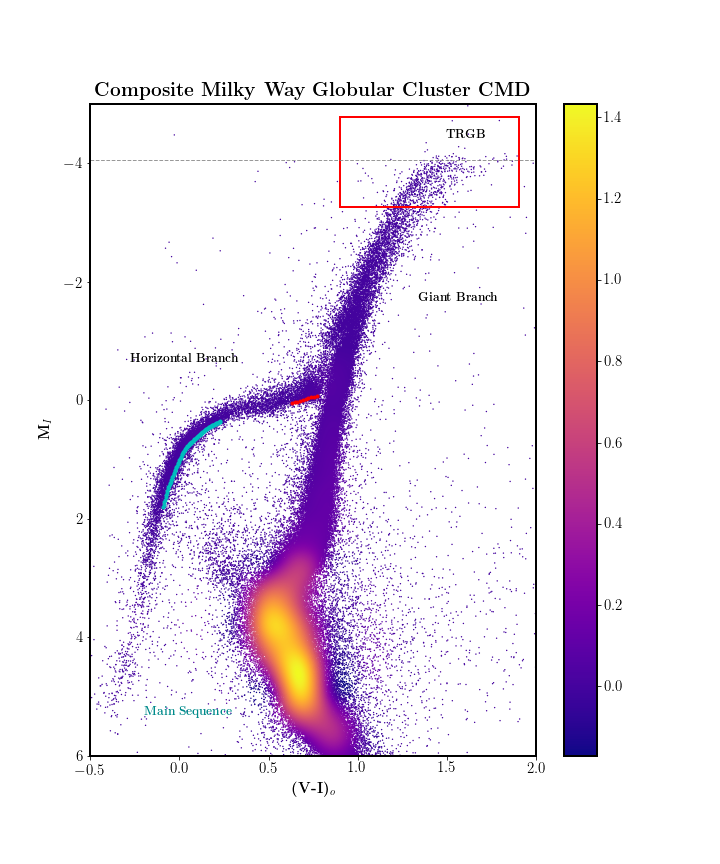}
     \caption{A composite M$_I$ versus $(V-I)_o$ color-magnitude diagram based on 46 Galactic globular clusters, color-coded by the density of points. The clusters span a range in metallicity of $-$2.4 $<$ [Fe/H] $< -$1.0 dex. Cluster membership was determined from their \gaia DR2 proper motions. The red rectangular box outlines the region of the red giant branch that is expanded in \autoref{fig:GC_TRGB}. The horizontal branch, main-sequence turnoff and giant branch are labeled. The horizontal gray dashed line indicates the TRGB at M$_I$ = -4.056 mag, and the cyan and red lines indicate the blue and red horizontal branch fits as measured by \citet{cerny_2021a}. }
\label{fig:GC_ALL}
\end{figure*}

A composite M$_I$ versus $(V-I)_o$ color-magnitude diagram (CMD) is shown in Figure \ref{fig:GC_ALL} for the 46 low-reddening clusters from \citet{cerny_2021a}. 
This composite shows a well-defined giant branch, sampling a wide range of metallicities from -2.4 $<$ [Fe/H] $<$ -1.0 dex. As described in more detail in \citeauthor{cerny_2021a}, high signal-to-noise and low-extinction clusters were used to define a fiducial lower envelope to the blue and red horizontal branches, and a maximum-likelihood grid search technique was used to align the remaining clusters onto a common calibration. The zero point of the calibration was set by the geometric DEB distance to $\omega$  Cen, measured by \citet{thompson_2001}. The resultant blue and red horizontal branches are shown in Figure \ref{fig:GC_ALL}.\footnote{Note that the process of aligning the clusters based on their horizontal branches is completely independent of the TRGB.}  Applying a Sobel edge-detection filter to the composite luminosity function for the TRGB, \citeauthor{cerny_2021a} determined an absolute $I$-band TRGB magnitude -4.056 mag, which, following F19, transforms to flight magnitudes as M$_{814W}$ = \trgbmaggc \ $\pm$ \trgbmaggcstat \ $\pm$ \trgbmaggcsys \ mag.

\subsubsection{\gaia Early Data Release 3 Calibration of Galactic Globular Clusters}
\label{sec:gaia_edr3}

With the ESA  \gaia mission,  the promise of astrometry reaching tens of microarcsecond accuracy \citep[Gaia Collaboration, ][]{prusti_gaia_2016} has been eagerly anticipated. Such astrometry for Galactic Cepheids, TRGB stars and other distance indicators will ultimately fix the absolute zero point of the extragalactic distance scale to an unprecedented accuracy of better than 1\%. However, in early data releases, it was discovered that there is a zero-point offset \citep[e.g.,][]{lindegren_2016}.  This offset results from the fact that the basic angle between the two \gaia telescopes is varying (resulting in a degeneracy with the absolute parallax). In addition, these variations lead to zero-point corrections that are a function of the magnitude, color, and position of the star on the sky \citep{lindegren_2018, arenou_2018}. In DR2, \citet{mignard_2018} and \citet{arenou_2018} found an average zero-point offset of -29 $\mu$as relative to the background reference frame for more than 550,000 quasars defined by the International Celestial Reference System.

Recently, the \gaia mission has released a new and updated database (Early Data Release 3; EDR3). This \gaia EDR3 database \citep{gaia_brown_2021} contains parallaxes, proper motions, positions and photometry  for 1.8 billion sources brighter than magnitude G=21 mag \citep{gaia_lindegren_2021b}.  The baseline for EDR3 is 34 months compared to 22 months for DR2, and thus provides a significant improvement to the astrometry.  The parallax improvement is estimated to be 20\% compared to DR2; in addition, the variance in the parallaxes (the systematic uncertainty), as measured over the sky and estimated from quasars, has been reduced by 30–40\% \citep{gaia_brown_2021}. Still, on average, the zero-point offset for EDR3 is found to be -17 $\mu$as (in the sense that the \gaia parallaxes are too small).  The \gaia collaboration has provided additional parallax  corrections for EDR3, which are again a function of G magnitude, color and ecliptic latitude \citep{gaia_lindegren_2021a}. 

However, as the \gaia Collaboration emphasizes \citep[e.g.,][]{bailer-jones_2021, fabricius_2021} there is a significant variance in these measured  offsets over the sky, and the EDR3 uncertainties in the parallaxes  for different objects are correlated as a function of their angular separations. \citet{ gaia_lindegren_2021a, gaia_lindegren_2021b} 
calculate the angular power spectrum of parallax systematic biases in \gaia EDR3 quasar data and estimate that the $rms$ variation of the parallax systematics (excluding the global offset) is about 10 $\mu$as on angular scales $>\sim$10 degrees. More recently,
\citet{maiz_apellaniz_2021} and \citet{vasiliev_baumgardt_2021} have analyzed  EDR3 parallax  data for a sample of Milky Way globular clusters. Both studies concur with the result that there are significant $rms$ variations on both   large  and small angular scales.  \citeauthor{maiz_apellaniz_2021}  conclude that the angular covariance limit results in  a {\it minimum} (and systematic) uncertainty for EDR3 parallaxes for individual stars or small-angular diameter clusters of 10.3 $\mu$as out to 30 arcmin. The $rms$ fluctuations can reach as high as 30-50 $\mu$as. They further note that the uncertainty cannot be significantly reduced for larger clusters.  

The minimum 10 $\mu$as systematic uncertainty in the EDR3 parallaxes limits the accuracy with which we can calibrate the TRGB for Galactic globular clusters.  \citet{cerny_2021a} (as described in \S \ref{sec:GC} above) based their calibration on the geometric DEB distance to  $\omega$  Cen, anticipating that in future, accurate Gaia parallax measurements for all 46 clusters will be available for calibration.    $\omega$  Cen has a measured Gaia EDR3 parallax of 189  $\mu$as or a distance of 5.25$^{+0.28}_{-0.25}$ kpc \citep{maiz_apellaniz_2021, vasiliev_baumgardt_2021}. Unfortunately,  a minimum systematic uncertainty of 10 $\mu$as results in a minimum (large) distance uncertainty of 5\% (0.1 mag) for $\omega$  Cen.  Additionally concerning, \citeauthor{vasiliev_baumgardt_2021} provide evidence that the \gaia distances are systematically (and significantly) smaller than the  previously published distances to these systems (the parallaxes are overestimated by 6-9 $\mu$as above the correction provided by \citet{gaia_lindegren_2021a}).

Based on \gaia EDR3 measurements for $\omega$  Cen, \citet{soltis_2021}  more optimistically quote a parallax measurement of  0.191 $\pm$ 0.001 (statistical) $\pm$ 0.004 (systematic) mas (2.2\% total uncertainty)  corresponding to a distance of 5.24 $\pm$ 0.11 kpc,  an uncertainty significantly smaller than (the minimum of 5\%) demonstrated by all of the studies discussed above. As \citet{vasiliev_baumgardt_2021} note, these $rms$
variations across the sky are irreducible at present and they thus conclude that the true uncertainty of the Soltis et al. result has been significantly underestimated.

Further independent constraints on the distance to $\omega$  Cen come from measurements of the RR Lyrae stars in the cluster. Recent near-infrared $JHK$ measurements by \citet{braga_2018}   result in  distances of 5.43-5.49 kpc (depending on their metallicity calibration) with quoted total uncertainties of 2\%, in good agreement with the DEB distance, as well as with a number of other published optical and near-infrared RR Lyrae measurements listed in their Table 8. To within the 1-$\sigma$ uncertainties, the recent RR Lyrae distance scale agrees with the \gaia EDR3 measurements of \citet{maiz_apellaniz_2021} and \citet{vasiliev_baumgardt_2021}.\footnote{More recently, \citet{baumgardt_vasiliev_2021} obtain a 1\%   distance to $\omega$  Cen by combining CMD fitting, RR Lyrae, DEBs, in addition to the new \gaia EDR3 distance, corrected for the systematic offset. They find a distance of 5.426 $\pm$ 0.047 kpc (their Table 2),  in excellent agreement with the results presented here.}

The  uncertainties (of order 5\%) in both the DEB and \gaia EDR3 distances for $\omega$  Cen are currently too large to provide the 1\% level of accuracy that will ultimately be required for a resolution of the tension in \ho.  For this paper, we adopt the \citet{cerny_2021a} calibration, with a distance of 5.44 kpc, and its (large) associated uncertainty of $\pm$5\% ($\pm$0.1 mag). As a result, it receives a lower weight in the determination of the value of \ho described in \S \ref{sec:hotrgb}. We note that adopting the \gaia EDR3 distance of 5.25 kpc with the same uncertainty of $\pm$5\%  increases \ho by only 0.1\% in the final analysis.

The \gaia parallaxes and additional measurements will continue to improve as longer time baselines are established over the course of the mission: the full potential of \gaia has yet to be realized. DR4 and DR5 are expected to be based on 5.5 and 10 years of data, respectively\footnote{https://www.cosmos.esa.int/web/gaia/science-performance}.

 \subsection{LMC Calibration}
\label{sec:LMC}

F20 measured the TRGB for the LMC using the OGLE “Shallow” survey data of \citet{ulaczyk_2012}\footnote{ The LMC data are available
at  http://www.astrouw.edu.pl/ogle/
ogle3/maps/}. In order to avoid
crowding/blending effects within the high-surface-brightness bar, the sample of stars analyzed was confined to stars outside of a circle of one degree
radius,  centered on the bar of the LMC. The LMC reddening and extinction were measured using VIJHK photometry, differentially with respect to two low-reddening galaxies, \ic and the SMC.
Based on the DEB \citep{pietrzynski_2019} distance modulus to the LMC of 
18.477 mag,   the extinction-corrected absolute magnitude of the TRGB for the $I$ band was found to be M$_I^{TRGB}$ =  -4.047 $\pm$ 0.022 (stat) $\pm$ 0.039 (sys) mag. The  \citeauthor{pietrzynski_2019} measurement is based on the  surface-brightness-color calibration for late-type giant stars, from which the angular diameters of giant stars can be measured to an accuracy of 0.8\%.  

Recently, \citet{hoyt_2021b} has undertaken a detailed remeasurement of the LMC TRGB based on OGLE-III photometry,  isolating regions where the edge-detection measurements are sharp and single-peaked. He illustrates that these same regions are also low in dust content, and located away from regions of star formation. He incorporates the new reddening and extinction maps of \citet{skowron_2021} determined from the colors of red clump stars based on OGLE-IV photometry. Adopting the 1\% distance to the LMC based on DEBs \citep{pietrzynski_2019}, he finds M$_I^{TRGB}$ =  -4.038 $\pm$ 0.012 (stat) $\pm$ 0.032 (sys) mag, consistent to within 1\% with the earlier results. A detailed description of the error budget and the adopted statistical and systematic uncertainties is given in his Table 3.
The systematic uncertainty includes a $\pm$0.01 mag term on the OGLE photometric zero point. An additional $\pm$0.01 mag systematic uncertainty is included in the ground-to-HST calibration  resulting in M$_{814}^{TRGB}$ = \trgbmaglmc $\pm$ \trgbmaglmcstat $\pm$ \trgbmaglmcsys ~mag. 

As an aside, we note that \citet{yuan_2019} argued that the F19 calibration of \ho based on the distance to the LMC  was in error. However, \citet{freedman_2020} and \citet{hoyt_2021b} describe in some detail a number of incorrect assumptions that were made by \citeauthor{yuan_2019} The excellent agreement found here between the completely independent LMC, \ngc 4258, SMC and Galactic globular-cluster calibrations argues even more strongly against the  claims made in \citeauthor{yuan_2019}   Moreover, even if the LMC were to be excluded from the TRGB calibration altogether, the resulting change in the overall value of \ho is insignificant ($<$1\%).

\subsection{The Small Magellanic Cloud (SMC)}
\label{sec:SMC}

 The interaction of the LMC and SMC has resulted in a tidally-extended structure to the SMC, which has historically complicated the measurement of the SMC distance. F20 measured the TRGB using published OGLE data\footnote{The SMC OGLE data are available at website http://www.astrouw.edu.pl/ogle/ogle3/maps/.} for the inner region of the SMC, thereby avoiding confusion with the more extended tidal tails.  They measured an  $I$-band magnitude for the TRGB of m$_I^{TRGB}$ = 14.93 mag,  adopting a foreground extinction value of A$_I$ = 0.056 mag.\footnote{The value quoted in \citet{freedman_2020} is for the extinction-corrected I$_o^{TRGB}$ and not for the apparent magnitude as stated. Adopting the distance modulus based on five previously-measured DEB measurements  (which yielded a value of $\mu_0$ = 18.965 mag), would result in a zero-point calibration for the TRGB of M$_I^{TRGB}$ = -4.035 $\pm$ 0.03 (stat) $\pm$ 0.05 (sys) mag.}   

Mapping out the inclined system with very high precision,  \citet{graczyk_2020} have recently measured a new DEB  distance to the central region of the SMC to an accuracy of better than 2\%,  based on the  surface-brightness-color calibration of \citep{pietrzynski_2019}.  Augmenting the sample of measured DEBs from their previously published sample (from 5 to 15, a three-fold increase), \citet{graczyk_2020}  determine a distance modulus of \smcdebwerr.
The SMC thus provides another opportunity for an updated and independent calibration of the TRGB. An advantage of the SMC is its low star formation rate and dust content.

\citet{hoyt_2021b}
 has also  undertaken an reanalysis of the SMC OGLE-III data incorporating the updated \citet{skowron_2021} reddening maps. He measures an apparent tip magnitude of  m$_I^{TRGB}$ = 14.93 mag.  A detailed description of the adopted statistical and systematic uncertainties is given in his Table 3.  Based on the new  \citet{graczyk_2020}  true DEB distance modulus he finds  M$_I^{TRGB}$ = -4.050 $\pm $ \trgbmagsmcstat\ (stat)  $\pm$ \trgbmagsmcsys \ (sys) mag, in excellent agreement with the \ngc~4258,  Milky Way globular-cluster, and the LMC calibrations discussed above. An additional $\pm$0.01 mag systematic uncertainty is included in the ground-to-HST calibration  resulting in M$_{814}^{TRGB}$ = \trgbmagsmc $\pm$ \trgbmagsmcstat $\pm$ \trgbmagsmcsys ~mag. 

\subsection{ Additional { Comparisons}}
\label{sec:othercalibs}

In the cases described in this section, we do not use these systems to calibrate \ho, but rather note their excellent consistency with the other calibrations presented here, lending further confidence to the overall calibration of the TRGB.

Two recent studies of the Sculptor (Tran et al. 2021, in prep.)  
and  Fornax (Oakes et al. 2021, in prep.)
dwarf spheroidal companions to the Milky Way provide additional calibrations of the TRGB, constituting consistency checks on the geometric calibrations (for the LMC, Milky Way, \ngc~4258 and the SMC) described above. Wide-field Magellan IMACS VI data were obtained for each galaxy, from which the position of the apparent TRGB, and the position of the horizontal branch were measured.  Tran et al. measured an extinction-corrected value of the apparent TRGB $I$-band magnitude for Sculptor of  m$^{TRGB}_{I_o}$ = 15.487 $\pm$ 0.057 $\pm$ 0.014 mag.  For Fornax, Oakes et al. found m$^{TRGB}_{I_o}$ = 16.75 $\pm$ 0.03 $\pm$ 0.01 mag.  Adopting the absolute calibration of the horizontal branch from \citet{cerny_2021a}, as described in \S \ref{sec:GC} above, and shown plotted in \autoref{fig:GC_ALL}, yields true distance moduli of  19.56 $\pm$ 0.03 $\pm$ 0.10 mag and 20.79 $\pm$ 0.02 $\pm$ 0.10 mag, for Sculptor and Fornax, respectively. These measurements yield absolute $I$-band calibrations of the TRGB (based on the horizontal branch) of -4.07 $\pm$ 0.06 $\pm$ 0.10 and -4.04 $\pm$ 0.04 $\pm$ 0.10 mag, again in excellent agreement with the independent calibrations based on \ngc~4258, the LMC and the SMC.

\begin{figure*} 
 \centering
\includegraphics[width=0.5\textwidth]{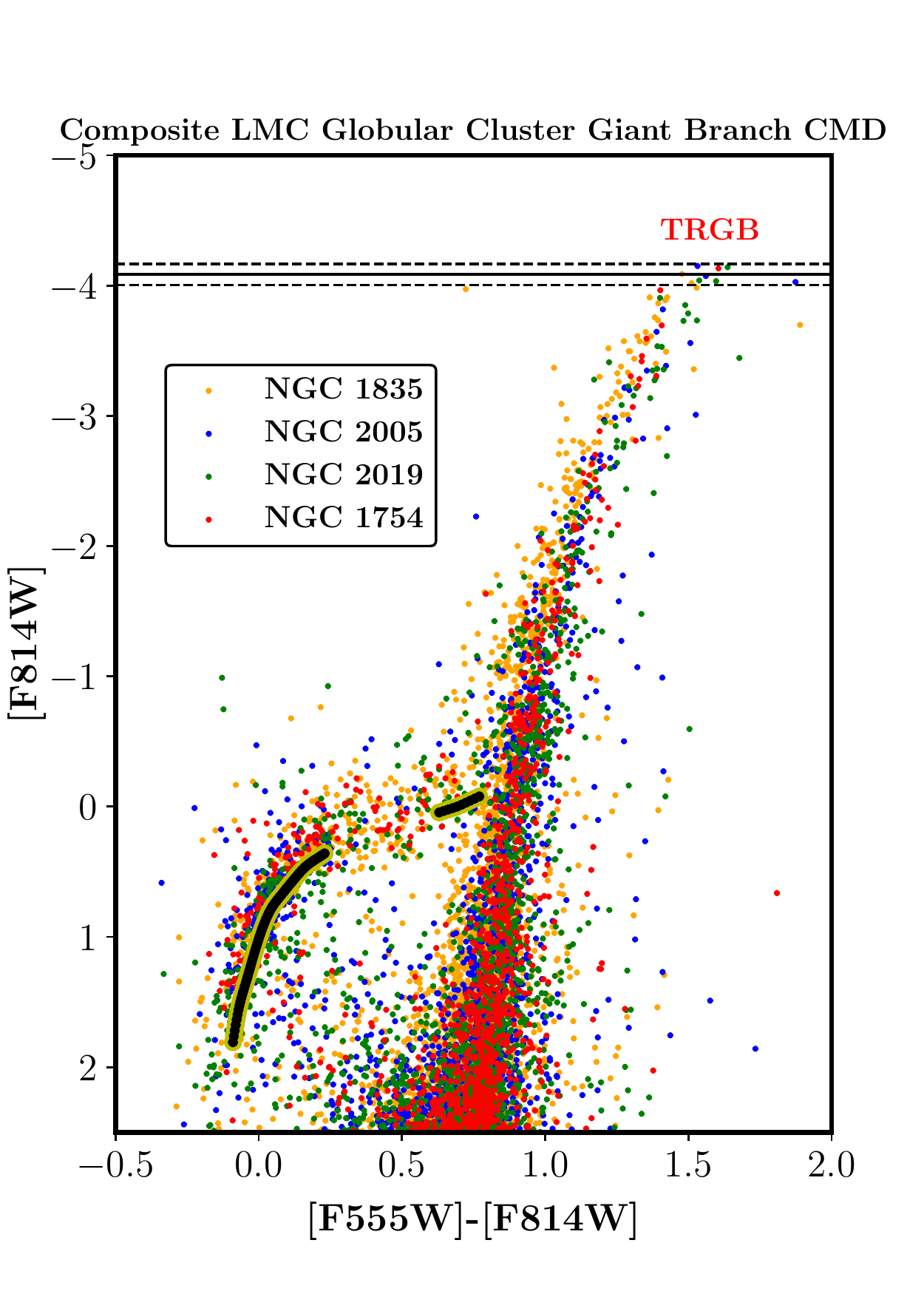}
     \caption{$I$ versus $(V-I)$ color-magnitude diagrams for four LMC globular clusters based on \hstacs data from  \citet{olsen_1998}. The blue and red fiducial horizontal branches defined by \citet{cerny_2021a} are shown.  The position of the tip and 1-$\sigma$ uncertainties are illustrated by the solid and dashed horizontal lines at the top of the figure. 
      }
\label{fig:lmcclustmontage}
\end{figure*}

Finally, we have also examined the $F814W $ and $F555W$ \hstacs data  obtained by \citet{olsen_1998} for a number of globular clusters in the LMC: specifically
 \ngc 1754, \ngc 1835, \ngc 2005, \ngc 2019. 
\autoref{tab:lmc_clusters} lists the reddenings and extinctions measured for each cluster by \citeauthor{olsen_1998}, and the true distance moduli based on the horizontal branch calibration of \citet{cerny_2021a}.
We show a composite color-magnitude diagram for these objects in \autoref{fig:lmcclustmontage}. Adopting the \citeauthor{cerny_2021a} calibration results in a measured TRGB magnitude of \trgbmaglmcclusters \ $\pm$ \trgbmaglmcclustersstat \ $\pm$ \trgbmaglmcclusterssys \ mag. 

\begin{deluxetable*}{llcc}
\tablecaption{Data for LMC clusters \label{tab:lmc_clusters}} 
\tablehead{\colhead{Cluster} & \colhead{$\mu_o$
} & \colhead{E(V-I)} & \colhead{A$_I$}} 
\startdata
NGC 2005  & 18.58   &   0.139 & 0.170 \\
NGC 2019  & 18.57   &   0.083 & 0.102 \\
NGC 1754  &  18.87  &   0.125 & 0.153 \\
NGC 1835  &  18.48  &   0.111 & 0.136 \\
\enddata
\end{deluxetable*}

As noted previously, these systems are not of comparable accuracy  (or independence) to yield an independent calibration of  \ho, but their consistency, to within the uncertainties, already provides a further test of the robustness of the TRGB calibration. In future, when parallaxes accurate to 1\% become available for a large sample of Milky Way globular clusters, these horizontal-branch measurements will become a powerful independent route to a calibration of the TRGB.

\vspace{1cm}
\subsection{Adopted TRGB Calibration}
\label{sec:trgbadopt}

\autoref{tab:trgbcal} lists the TRGB absolute magnitude at $F814W$ for the geometric calibrations described above. Where the calibration was carried out for ground-based data (as for the LMC, SMC and the Milky Way clusters), these have been transformed to the \hstacs $F814W$ flight-magnitude system. The \ngc~4258 calibration was carried out entirely with \hst and is already on the F814W flight magnitude system. As discussed in F19 and F20, the transformation from the $I$-band to $F814W$ results in a zero point that is brighter by -0.0068~mag.   As can be seen from this table, the good agreement of the TRGB zero point based on the calibrations for many anchors means that the adoption or rejection of a particular galaxy does not significantly impact the overall result.

\begin{deluxetable}{lllll} 
\tabletypesize{\normalsize}
\setlength{\tabcolsep}{0.05in}
\tablecaption{TRGB zero-point calibration 
\label{tab:trgbcal}}
\tablewidth{0pt}
\tablehead{ \colhead{Object}  &  \colhead{M$_{F814W}^{TRGB}$ (mag)} & \colhead{$\sigma_{stat}$} & \colhead{$\sigma_{sys}$}  & \colhead{Reference}}
\startdata
\ngc 4258  \tablenotemark{a}       & \trgbmagmaser  & \trgbmagmaserstat & \trgbmagmasersys            & \citet{jang_2021} \\
            &   &  &          \\
Milky Way globular clusters \tablenotemark{b}   & \trgbmaggc\tablenotemark{c}  & \trgbmaggcstat & \trgbmaggcsys &    \citet{cerny_2021a} \\
                   &   &  & \\
LMC       \tablenotemark{d}                   & \trgbmaglmc\tablenotemark{c}  & \trgbmaglmcstat &  \trgbmaglmcsys &  \citet{hoyt_2021b} \\
                        &   &  &  \\ 
SMC       \tablenotemark{e}                   & \trgbmagsmc\tablenotemark{c} & \trgbmagsmcstat & \trgbmagsmcsys &  \citet{hoyt_2021b} \\
\hline
Sculptor       \tablenotemark{f}                   & \trgbmagsculptor\tablenotemark{c} & \trgbmagsculptorstat & \trgbmagsculptorsys &  Tran et al. (2021) \\
Fornax       \tablenotemark{g}                   & \trgbmagfornax\tablenotemark{c} & \trgbmagfornaxstat & \trgbmagfornaxsys &  Oakes et al. (2021) \\
LMC globular clusters       \tablenotemark{h}                   & \trgbmaglmcclusters\tablenotemark{} & \trgbmaglmcclustersstat & \trgbmaglmcclusterssys &  this paper \\
\hline
Adopted Value
(MW, NGC 4258, LMC,   SMC)               & \trgbmagnew & \trgbmagnewstat &  \trgbmagnewsys  & this paper (\S \ref{sec:trgbadopt}) \\
\hline \hline
\enddata
\tablenotetext{a}{~~H$_2$O Megamaser distance calibration}
\tablenotetext{b}{~~Optical data, Gaia proper motion selection; $\omega$~Cen DEB calibration;  M$_I$ = -4.056 mag. }
\tablenotetext{c}{~~Transformation to M$_{F814W}^{TRGB}$ = M$_I$ - 0.0068 mag following \citet{freedman_2019}.}
\tablenotetext{d}{~~LMC DEB calibration; M$_I$ = -4.038 mag. An additional $\pm$ systematic uncertainty is included in the ground-to-HST calibration. }
\tablenotetext{e}{~~SMC DEB calibration ; M$_I$ = -4.050 mag. An additional $\pm$ systematic uncertainty is included in the ground-to-HST calibration.}
\tablenotetext{f}{~~$\omega$ Cen DEB calibration ; M$_I$ = -4.07 mag}
\tablenotetext{g}{~~$\omega$ Cen DEB calibration ; M$_I$ = -4.04 mag}
\tablenotetext{h}{~~$\omega$ Cen DEB calibration }
\end{deluxetable} 

\begin{figure*} 
 \centering
\includegraphics[width=0.71\textwidth]{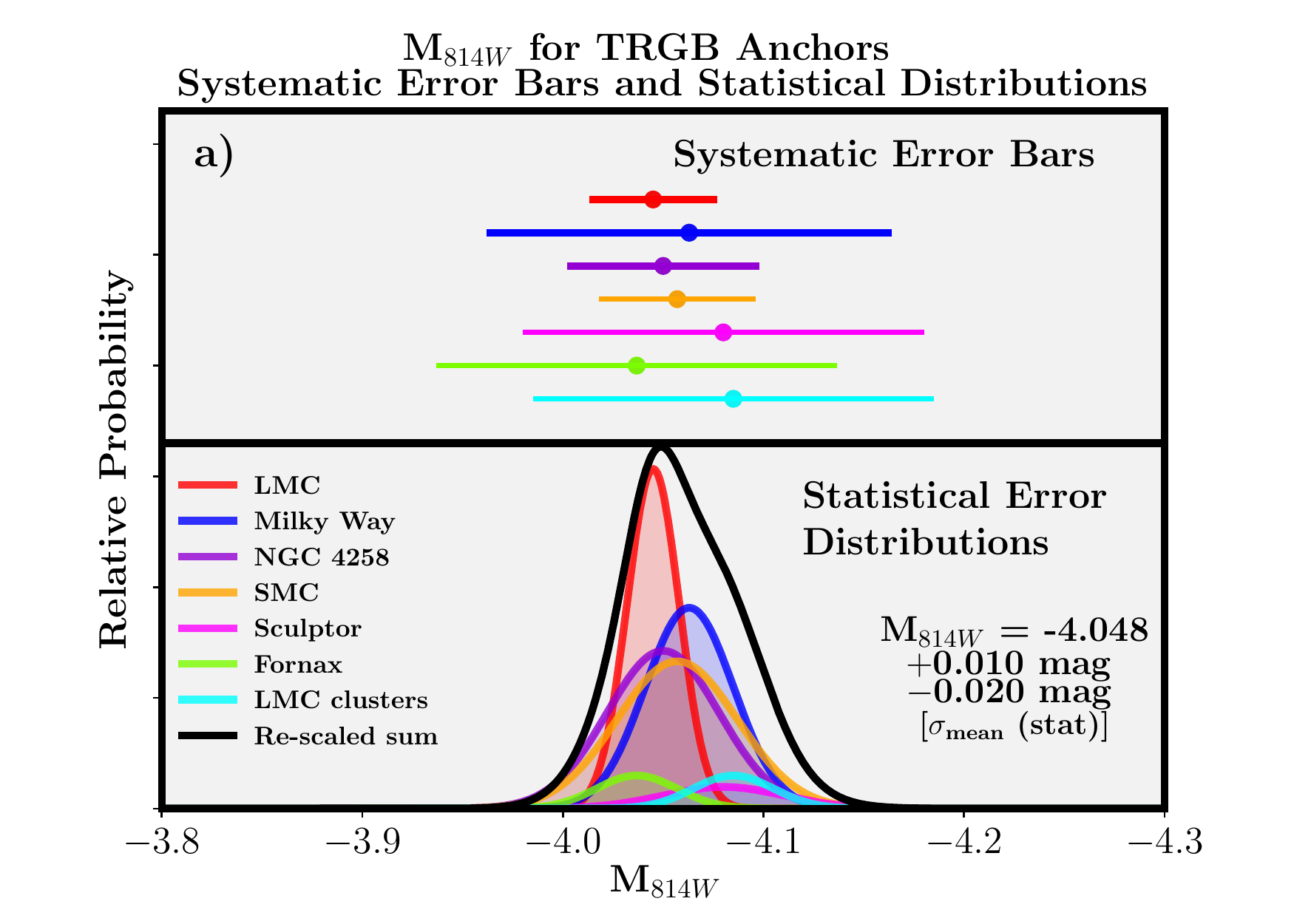}
\includegraphics[width=0.71\textwidth]{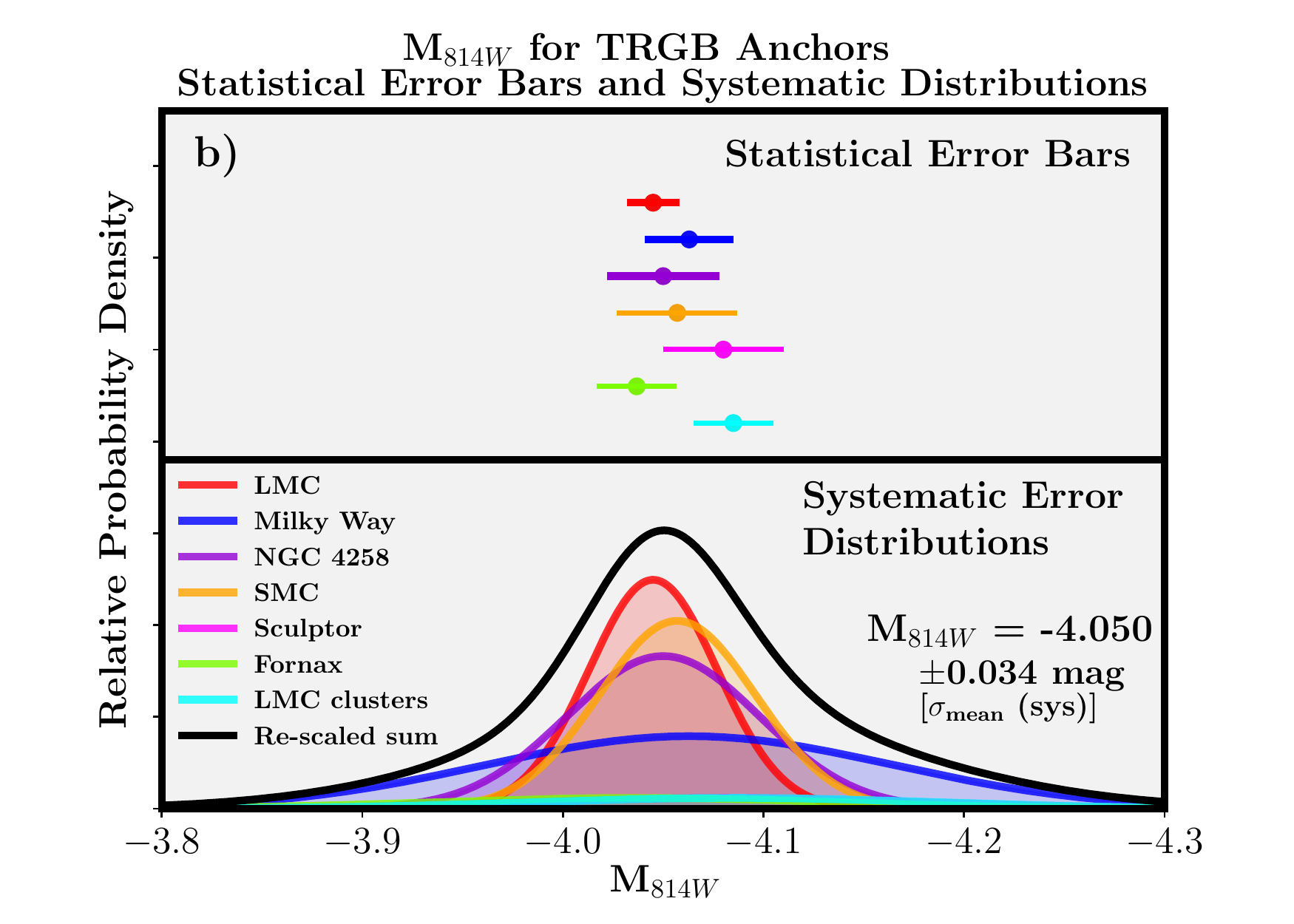}
     \caption{Probability density functions for the measured absolute magnitude of the TRGB. 
     The statistical and systematic errors are shown separately, so that the relative contributions of each can be easily seen for each galaxy. The statistical uncertainties can be improved by increasing the sample size in future,  decreasing as 1 / $\sqrt(N)$. In \autoref{fig:absmagtrgbPDFs}a), the systematic error bars are shown at the top of the plot, and the statistical error distributions are shown at the bottom. In \autoref{fig:absmagtrgbPDFs}b), the statistical error bars are shown at the top, and the systematic error distributions are shown at the bottom. As discussed  in \S \ref{sec:overall_systematics}, there is some covariance in the systematic uncertainties. Shown are the sum of all of the PDFs (black), the LMC (red), Milky Way (blue), \ngc~4258  (purple), SMC (orange), Sculptor (magenta), Fornax (green) and the composite of the four LMC globular clusters (cyan). The results for Sculptor, Fornax and the LMC clusters are shown for comparison purposes only. The statistical and systematic errors on the mean are labeled, along with the adopted value of \trgbmagnewlong, {consistent, to within 0.001 mag}, with the mode of the summed distribution in each case.}
\label{fig:absmagtrgbPDFs}
\end{figure*}

\autoref{fig:absmagtrgbPDFs} shows the relative probability density functions (PDFs) for the absolute TRGB $F814W$ magnitudes discussed in \S \ref{sec:trgbupdate} above. Here we separate the contributions of the statistical and systematic errors in each case, so that the contribution of both types of uncertainties can be clearly seen. In \autoref{fig:absmagtrgbPDFs}a), the widths of the Gaussians represent the individual statistical errors in each determination only, whereas the systematic uncertainties are illustrated separately by the error bars at the top of the plot (using the same color coding) for each object. Conversely, in \autoref{fig:absmagtrgbPDFs}b), the widths of the PDFs represent the individual systematic errors in each determination only, whereas the statistical  uncertainties are illustrated separately by the error bars at the top of the plot. The integrals of the PDFs for the LMC, Milky Way, \ngc~4258 and the SMC each have unit area. The statistical and systematic errors {\it for each individual determination}, $\sigma_i$, are given by the 16th and 84th percentiles of the Gaussians in Figures \ref{fig:absmagtrgbPDFs}a and b, respectively. The Frequentist sums of the probability distributions are shown in both cases by the black lines.  For the total sample, $\sigma_{\bf mean}$ =  $\sum{\sigma_{\bf i}} / \sqrt{(N-1)}$, where N = 4. 

As we have seen, the Milky Way TRGB magnitude is based on a sample of 46 clusters calibrated to the DEB distance to $\omega$ Cen. The calibrations of Sculptor, Fornax and the LMC clusters are not independent, however, since they all rely on the Milky Way calibration of the horizontal branch. Moreover, Sculptor and Fornax are single objects, and the TRGB for the LMC clusters is sparsely populated. For illustrative purposes, in Figures \ref{fig:absmagtrgbPDFs}a) and b), the areas for these Gaussians have thus been down-weighted by a factor {\it 1/f} as shown in \autoref{eq:scalearea}:
\begin{equation}
\label{eq:scalearea}
    {1\over{f \sqrt{2\pi\sigma^2}  } } 
e^{-0.5 \bigl({ x-<x>\over{\sigma^2}}\bigr)} 
\end{equation}
\noindent
where {\it f} = $\sqrt{(N)}$, and N = 46, the size of the Milky Way globular cluster sample. Thus Sculptor, Fornax and the LMC clusters do not contribute to the adopted overall calibration, but they do provide a consistency check on the horizontal branch to TRGB distance scale. 

\begin{figure*} 
 \centering
\includegraphics[width=0.71\textwidth]{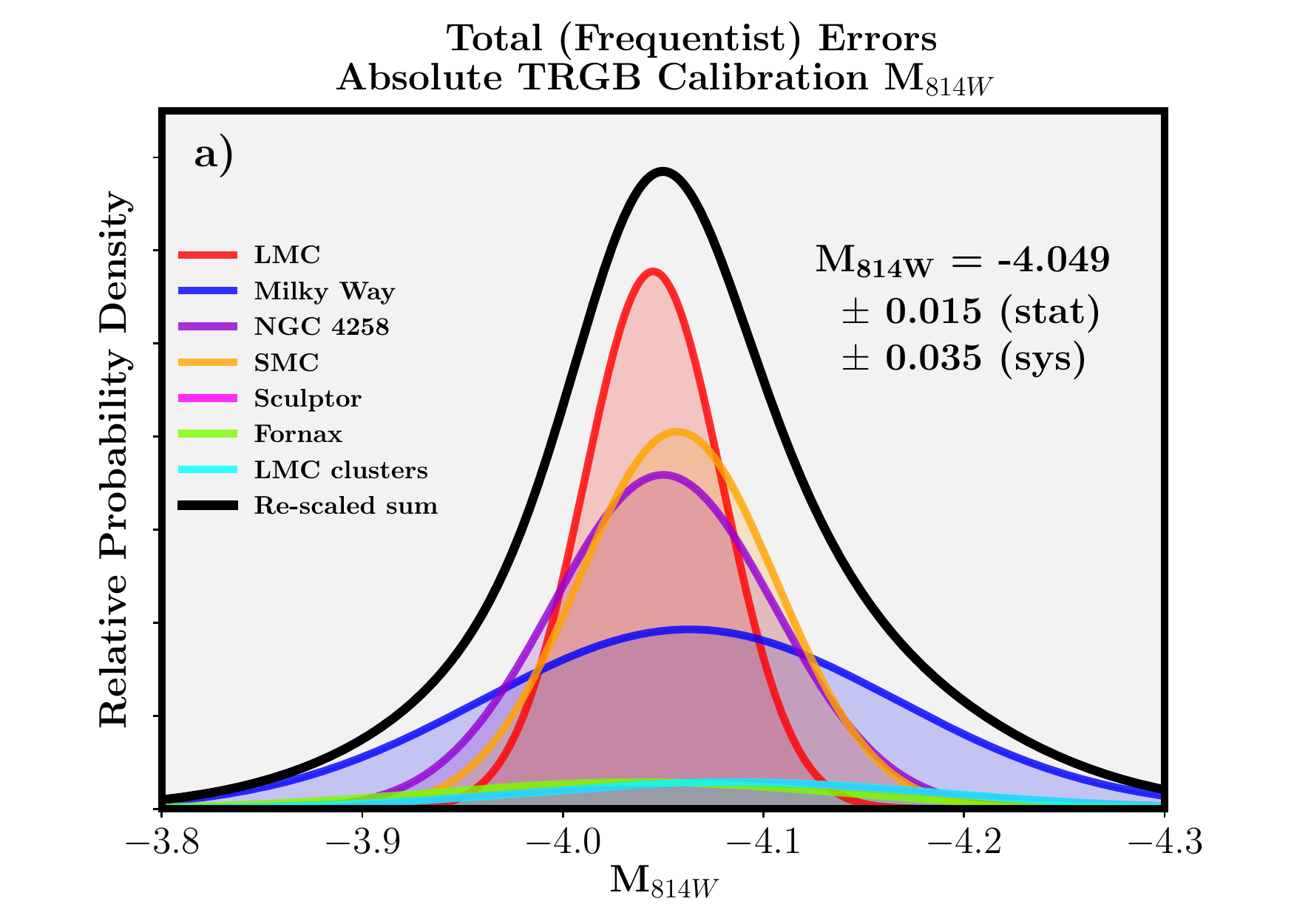}
\includegraphics[width=0.71\textwidth]{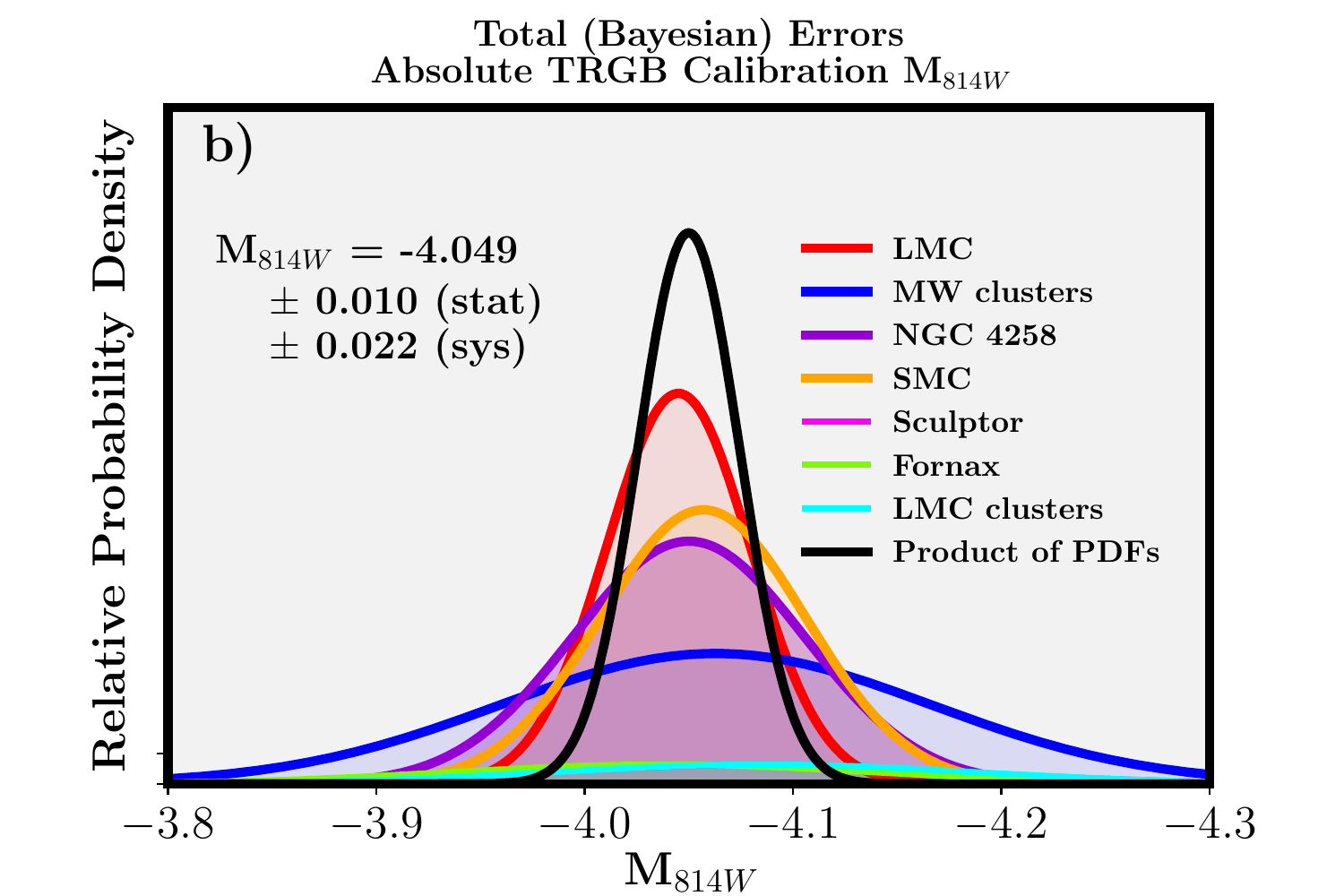}
     \caption{Probability density functions for the measured absolute magnitude of the TRGB. The  total errors (statistical and systematic, combined in quadrature) are shown, as described in the text.  \autoref{fig:absmagtrgbPDFs_tot}a) The Frequentist sum of all of the PDFs (black), the LMC (red), Milky Way (blue), \ngc~4258  (purple), SMC (orange), Sculptor (magenta), Fornax (green) and the composite of the four LMC globular clusters (cyan). The statistical and systematic errors on the mean are labeled, as described in the text, along with the adopted value of 
     \trgbmagnewlong. \autoref{fig:absmagtrgbPDFs_tot}b) The product of the PDFs. Color scheme is the same as that for Figure \ref{fig:absmagtrgbPDFs}.}
\label{fig:absmagtrgbPDFs_tot}
\end{figure*}

The Frequentist sums of the probability distributions are shown  by the black lines in \autoref{fig:absmagtrgbPDFs_tot}a. The  mode of the summed distribution for the four primary TRGB calibrators   is \trgbmagnew \ mag.  As shown in  \autoref{fig:absmagtrgbPDFs_tot}b, an identical result is obtained for a Bayesian analysis (albeit with smaller uncertainty), in which a uniform prior is adopted, and the product of the distributions is determined. In addition, a simple weighted average for the LMC, Milky Way, \ngc~4258 and the SMC also gives a result to within 0.001 mag of \trgbmagnew \ mag. We adopt this robust value, \trgbmagnewlong,  for the absolute magnitude of the TRGB.  The (exact) agreement of the various means of combining the four calibrations lends confidence to the overall result; i.e., it is independent of the choice of statistical approach adopted to combine the results. Finally, we note that this value agrees to better than 1\% with that given by F20, who found M$^{TRGB}_{F814W}$ = \absmagFb \ $\pm$ \absmagstatFb \ $\pm$ \absmagsysFb \ mag.
\vspace{1cm}

\section{The Hubble Constant Based on the TRGB}
\label{sec:hotrgb}

\subsection{New TRGB Calibration of H$_o$ Based on Supernovae Ia}
\label{sec:newtrgbho}

We turn now to a determination of \ho based on the TRGB calibration discussed in
\S\ref{sec:trgbadopt}. 
This is an update of the calibration of \ho presented in F19 and F20, applied to the Carnegie Supernova Project (CSP) sample of 99 \sne observed at high cadence and multiple wavelengths \citep{krisciunas_2017}. That measurement of \ho was based on \hstacs observations of the halos of 15 galaxies that were hosts to 18 \sne. The measured absolute $I$-band magnitude of the TRGB from \citet{freedman_2020} was M$_{F814W}$ = -4.054 $\pm$ 0.022 (stat) $\pm$ 0.039 (sys) mag, tied to  the geometric DEB distance modulus to the LMC from \citet{pietrzynski_2019} of 18.477 mag. We now use four independent calibrations (\ngc~4258, Milky Way, LMC and SMC) superseding the single calibration based on the LMC alone. 

To briefly summarize, in F19 the  CSP   analysis was undertaken with the SNooPy package \citep{burns_2018}, which characterizes the \sne light-curve shape using a color-stretch parameter, s$_{BV}$. Magnitudes were computed using two different approaches to the reddening where
\begin{equation}
\label{eq:CSPmags}
B^\prime  =  B - P^1(s_{BV}-1) - P^{2}(s_{BV}-1)^2 - CT - 
       \alpha_M(\log_{10}M_*/M_\odot-M_0),
\end{equation}
\noindent
where $P^1$ is the linear coefficient and $P^2$ is the quadratic coefficient in ($s_{BV}-1$);  $B$ and $V$ are the apparent, K-corrected peak
magnitudes;  $\alpha_M$ is the slope of the correlation between peak luminosity
and host stellar mass $M_*$; and CT denotes the color term for the two approaches. In the first case,  CT = $\beta$(B-V), where a color coefficient $\beta$ results in a reddening-free magnitude, an approach originally proposed by \citet{tripp_1998}. In the second approach, CT = R$_B$ E(B-V), where R$_B$, the ratio of total-to-selective absorption, and the reddening, E(B-V), are solved for explicitly using both optical and near-infrared  colors for each \sn.    Using the MCMC fitter described in \citet{burns_2018} and F19,
which uses the “No U-Turn Sampler” from the data modeling language STAN \citep{carpenter_2017},
and solving for the parameters in \autoref{eq:CSPmags}, the value of \ho and its error were obtained using both approaches. 

As described in \citet{hoyt_2021a}, two new galaxies with directly measured TRGB distances have been added to the \cchp sample since F19 and F20. \ngc~5643 is host to SN 2013aa
and SN 2017cbv \citep{burns_2020}; and \ngc~1404,  a member of the Fornax cluster, is host to SN 2007on and SN 2011iv \citep{gall_2018}.  \citet{hoyt_2021a} find that SN 2007on appears to be significantly underluminous, and it is therefore excluded from the current analysis. In F19, the distance to \ngc~1404 was taken to be the average value given by the two other Fornax galaxies in the \cchp sample, \ngc~1316 and \ngc~1365.  In this paper, we adopt the new direct distance to \ngc~1404 \citep{hoyt_2021a} for SN 2011iv and add the two additional \sne in \ngc~5643,  augmenting the sample of 18 \sne described in F19 to an updated sample of 19. All distances are calibrated adopting \trgbmagnewlong, and used as new input to the MCMC analysis described in F19  (C. Burns, priv. comm.). 

\begin{deluxetable*}{c|cccc|cccc|c}
\tablewidth{0pc}
\tablecaption{Values of $H_0$  
             $\mathrm{(km\ s^{-1} Mpc^{-1})}$ for various choices of fit  \label{tab:CSP_ho}}
\tablehead{
 & \multicolumn{4}{c|}{Tripp} & 
\multicolumn{4}{c|}{$E(B-V)$} & \\
Band & $H_0$ (CSP18)\tablenotemark{a} & $\sigma$ & $H_0$ (CSP20)\tablenotemark{b} & $\sigma$ &
          $H_0$ (CSP18)\tablenotemark{a} & $\sigma$ & $H_0$ (CSP20)\tablenotemark{b} & $\sigma$ }
\startdata
\multicolumn{10}{c}{Full Sample} \\
$B$ & $69.48 $ & $1.39 $ & $69.88$ & $1.25 $ & $70.75$ & $1.32 $ & $71.50$ & $1.29 $  \\
$H$ & $69.13$ & $1.35 $ & $70.48 $ & $1.23 $ & $69.36$ & $1.46 $ & $70.33$ & $1.41 $  \\
\multicolumn{10}{c}{$s_{BV} > 0.5$ and $(B-V) < 0.5$} \\
$B$ & $69.38 $ & $1.36 $ & $69.57 $ & $1.24 $ & $69.39 $ & $1.04 $ & $70.04$ & $1.05 $  \\
$H$ & $68.80 $ & $1.34 $ & $70.00 $ & $1.25 $ & $68.47 $ & $1.44 $ & $69.47 $ & $1.42 $  \\
\enddata
\tablenotetext{a}{~CSP \sne host-galaxy-mass corrections from \citet{burns_2018}.  }
\tablenotetext{b}{~CSP \sne host-galaxy-mass corrections from \citet{uddin_2020}.  }
\end{deluxetable*}

\begin{figure*} 
 \centering
\includegraphics[width=1.0\textwidth]{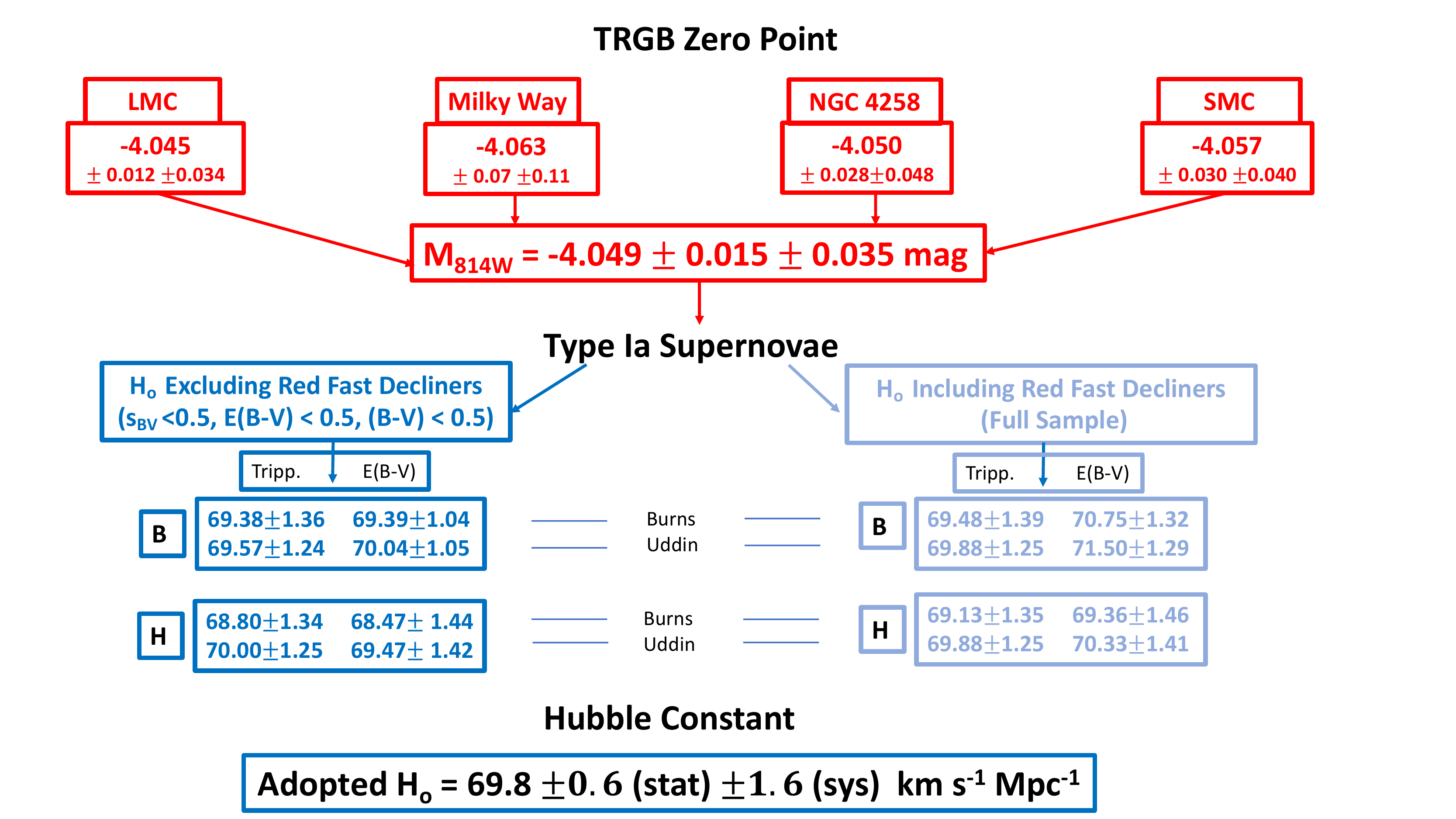}
     \caption{An overall flowchart summarizing the results of the TRGB zero-point calibration described in \S \ref{sec:trgbupdate} and the \sne calibration described in \S \ref{sec:hotrgb}, leading to the adopted value of \honewlongnospace. The TRGB zero-point is based on the M$_{814W}$ calibrations for the LMC, SMC, \ngc~4258 and the Milky Way. The adopted value of \ho is based on a sample of \sne restricted to those with $s_{BV} > $ 0.5 and (B - V) $<$ 0.5, for which there is good proportional overlap between the TRGB and more distant host galaxy samples.
\label{fig:hoflowchart}}
\end{figure*}

In \autoref{tab:CSP_ho}, we give the values of \ho and uncertainties obtained adopting the new calibration of M$^{TRGB}_{F814W}$. Listed are \ho values based on both the  CSP $B$-band and $H$-band \sne magnitudes for different color and dust reddening constraints. The uncertainties (for the \sne analysis alone) are determined from a diagonal covariance matrix with respect to the TRGB distances. Following F19, we present results applying both the Tripp and explicit E(B-V) reddening corrections. In addition, we present the results adopting host-galaxy masses from \citet{burns_2018} originally used in F19, as well as those measured in a more recent study of \citet{uddin_2020}. \autoref{fig:hoflowchart} shows these results in flowchart form.

\begin{figure*} 
 \centering
\includegraphics[width=0.8\textwidth]{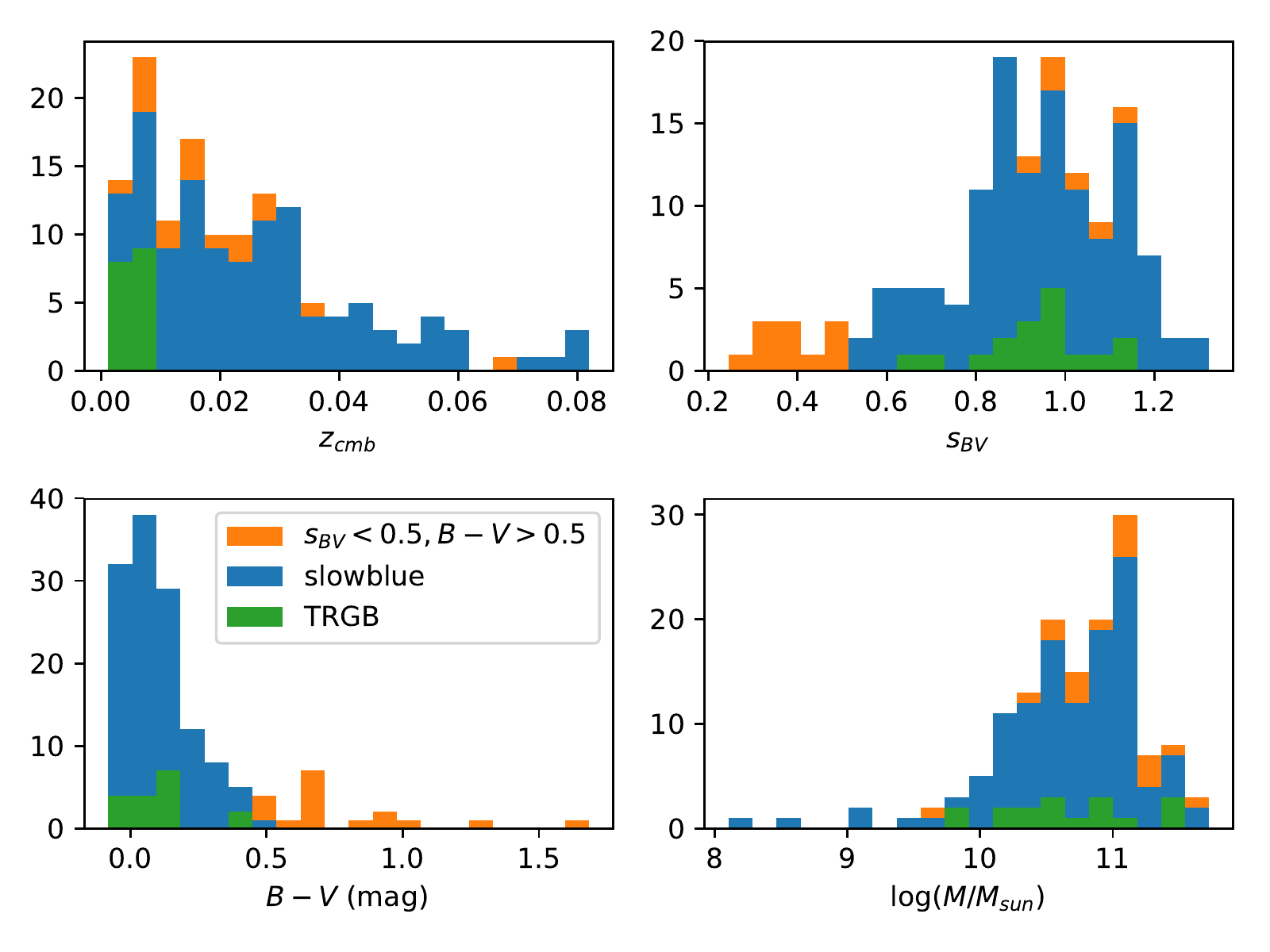}
     \caption{The upper left panel shows the redshift distribution for the total sample of CSP \sne. In blue are the slow decliners (with $s_{BV} > $ 0.5 and (B - V) $<$ 0.5), labeled ``slowblue". In orange are the red, fast decliners, and the nearby calibrating galaxies with measured TRGB distances are shown in green. The upper right panel shows the distribution of stretch values, and the lower two panels show the distributions of (B-V) and log$_{10}$($\mathrm M \over M_{\odot}$), respectively. The ``Full Sample" in \autoref{tab:CSP_ho} includes both the orange and blue distributions (i.e., the different samples are not overplotted, and no orange bins are being lost). The green TRGB distribution is well-matched to that of the of slower, bluer decliners, and does not exhibit the extended tails seen in orange for stretch (with s$_{BV} < 0.5$) and color ($(B-V) > 0.5$) of the red, fast decliners.}
\label{fig:SNehists}
\end{figure*}

The values presented in \autoref{tab:CSP_ho} and \autoref{fig:hoflowchart} represent different choices for: the \sn sample (color and stretch); dealing with dust (Tripp versus E(B-V)); bandpass for the \sn magnitudes ($B$ versus $H$);
and  host galaxy-mass peak \sn luminosity correlation (Burns versus Uddin).  The various choices result in a full range in \ho values from 68.47 to 71.50 \hounits.  
In selecting a best value of \ho from those listed in \autoref{tab:CSP_ho}, we select  (following F19) the sample that minimizes the difference between the calibrator sample and the distant sample in terms of the nuisance variables:  color, stretch, and host mass. In the histograms in \autoref{fig:SNehists} we illustrate the characteristics of the  \sn in the distant galaxy sample compared with those for the TRGB calibrators. The TRGB sample is shown in green; the overall CSP sample is divided such that the blue, slow decliners (with $s_{BV} > $ 0.5 and (B - V) $<$ 0.5) are shown in blue (and labeled ``slowblue") and those with fast decline rates and redder colors are shown in orange.  In terms of stretch and color, the tails seen in orange (extending to s$_{BV} <$ 0.5 and ($B-V) > 0.5$) are absent in the TRGB calibrating sample. Unambiguously, in terms of stretch and color, the sample with the best  overlap of calibrator and distant \sne is that of “slowblue”.   This sample also overlaps well in terms of host-galaxy  mass, an advantage of the TRGB method, which can be applied to both early- and late-type galaxies. (Cepheid
variables are young objects found only in star-forming (e.g., spiral)
galaxies and cannot calibrate the \sne found in elliptical or S0
galaxies.)

We note the following:
\begin{enumerate}
    \item Using B-band photometry, restricting the sample to that with $s_{BV} > $ 0.5 and (B - V) $<$ 0.5 (``slowblue"), and basing the analysis on the more recent \citet{uddin_2020} host-galaxy masses results in a value of \ho = 69.57 $\pm$ 1.24 \hounits using the Tripp  method and \ho = 70.04 $\pm$ 1.05 \hounits  explicitly correcting for dust (E(B-V)). Using H-band photometry, respectively results in similar values of  \ho = 70.00 $\pm$ 1.25  and \ho = 69.47 $\pm$ 1.42  \hounits. The corresponding values based on the \cite{burns_2018} masses are slightly lower. The difference arises primarily because the slope of the mass correlation in the optical is steeper for the Burns masses, whereas for the Uddin masses, the relation is nearly flat for all filters. 
    \item The H-band data have the advantage of smaller dependence on the reddening, as the correction (R$_\lambda$) is smaller, but they have the disadvantage of  larger variance because the sample of \sne having H-band photometry is smaller. (The ``Full Sample" has 147 objects; ``slowblue" restricts the sample to 129 objects; and restricting the sample to those with H-band photometry results in 102 objects.) 
    \item The largest value of \ho (71.5) is obtained when the redder, faster decliners are included in the analysis (the ``Full Sample"). However, as noted above, these solutions are strongly disfavored since there are no redder, faster decliners in the more distant sample. In a broader context,  no solution here reaches a value as high as 74 \hounits. 
\end{enumerate}
    
Although the differences in these \ho values are small (a total range of only 3 \hounits), they illustrate the effect of different choices  in the host-galaxy mass correlation and method/filters adopted to correct for dust. 

Our adopted best-fit value  is based on 1) the sample of \sne for which the nuisance variables  (color, stretch, and host mass) are comparable for the calibrating TRGB galaxies and the distant \sne; 2) an average of the Tripp/E(B-V) determinations; 3) the recent host-galaxy masses measured by \citet{uddin_2020}. We choose the B-band measurements because the sample of \sne is largest, and the scatter for the H-band measurements is 40\% larger (or a factor of two in the variance, in the case of the E(B-V) correction).  We adopt a best-fit value of 69.8 $\pm$ 1.2 (sys) \hounits. This latter uncertainty takes into account the systematic uncertainties in the \sn analysis alone, without yet combining it with the TRGB zero-point systematic error. As discussed in \citet{burns_2018,freedman_2019}, all of the correction factors to the \sn light curves ($P^1$, $P^2$, $s_{BV}-1$, $\alpha_M$, $\beta$, E$(B-V)$, R$_B$), as described in \S \ref{eq:CSPmags} are computed; these then provide corrected magnitudes and a full covariance matrix, used to determine \ho and the uncertainty given in \autoref{tab:CSP_ho}. The total uncertainty adopted for \ho, including the uncertainty in the TRGB calibration is discussed below. 

The \ho values and uncertainties based individually on the  new  TRGB calibrations for  \ngc 4258 (\S  \ref{sec:n4258}), Galactic globular clusters (\S \ref{sec:GC}),  the SMC (\S \ref{sec:SMC}), and the LMC  \S \ref{sec:LMC} 
are listed in  \autoref{tab:hotrgbceph}.   For comparison, also listed are the  \ho values, their uncertainties and their published references from the SHoES team, based on the Cepheid calibrations for the LMC, \ngc 4258 and the Milky Way. 

Both statistical and systematic uncertainties are given for the TRGB \ho determinations in \autoref{tab:hotrgbceph}. The error bars include both the uncertainties for the TRGB calibration discussed in \S \ref{sec:trgbadopt} above, in addition to those arising from the calibration of the \sne, as discussed above, and in F19. For the \sne, the  statistical uncertainty amounts to $\pm$ 0.5\% with a systematic uncertainty  of $\pm$1.7\%. The final percentage errors are summarized in Table \ref{tab:trgbhoerrors}, with a  final adopted value of \honewlongnospace.

\begin{deluxetable}{lccl} 
\tabletypesize{\normalsize}
\setlength{\tabcolsep}{0.05in}
\tablecaption{H$_0$ Values for Common TRGB and Cepheid Calibrators
\label{tab:hotrgbceph}}
\tablewidth{0pt}
\tablehead{ \colhead{Calibrator}  &   \colhead{H$_0$ (TRGB) } & \colhead{H$_0$ (Cepheids)\tablenotemark{a}} & \colhead{Cepheid Reference}}
\startdata
LMC        &  69.9 $\pm$ 0.5 (stat) $\pm$ 1.6 (sys) &  74.22 $\pm$ 1.82 & \citet{riess_2019} \\
\ngc 4258  &  69.7 $\pm$ 1.0 (stat) $\pm$ 2.0 (sys) & 72.0 $\pm$ 1.9 & \citet{reid_2019} \\
Milky Way  &  69.3 $\pm$ 0.8 (stat) $\pm$ 3.5 (sys) & 73.0 $\pm$ 1.4 & \citet{riess_2021}\\
SMC        &  69.5 $\pm$ 1.0  (stat) $\pm$ 1.7 (sys) &  ... & ... \\
\hline 
Adopted Value  &  \honew \ $\pm$ \honewstaterr \ (stat) $\pm$ \honewsyserr \ (sys) & 73.2 $\pm$ 1.3 & \citet{riess_2021} \\
\hline \hline
\enddata
\tablenotetext{a}{~The published SHoES \ho results are given with total errors only.}
\end{deluxetable} 

\begin{deluxetable*}{lccc}
\tablecaption{Summary of \ho Uncertainties \label{tab:trgbhoerrors}} 
\tablehead{\colhead{Source of Error} & \colhead{Random Error} & \colhead{Systematic Error} & \colhead{Description}} 
\startdata
TRGB Zero Point       &       0.7\%       &        1.6\%      &      \S \ref{sec:trgbadopt}  \\
\cspi \sne &       0.5\%       &        1.7\%      &   F19, \S \ref{sec:hotrgb}   \\         
 \hline
Total &             0.9\%       &        2.3\%   &  In quadrature \\
\enddata
\end{deluxetable*}

\begin{figure*} 
 \centering
\includegraphics[width=1.0\textwidth]{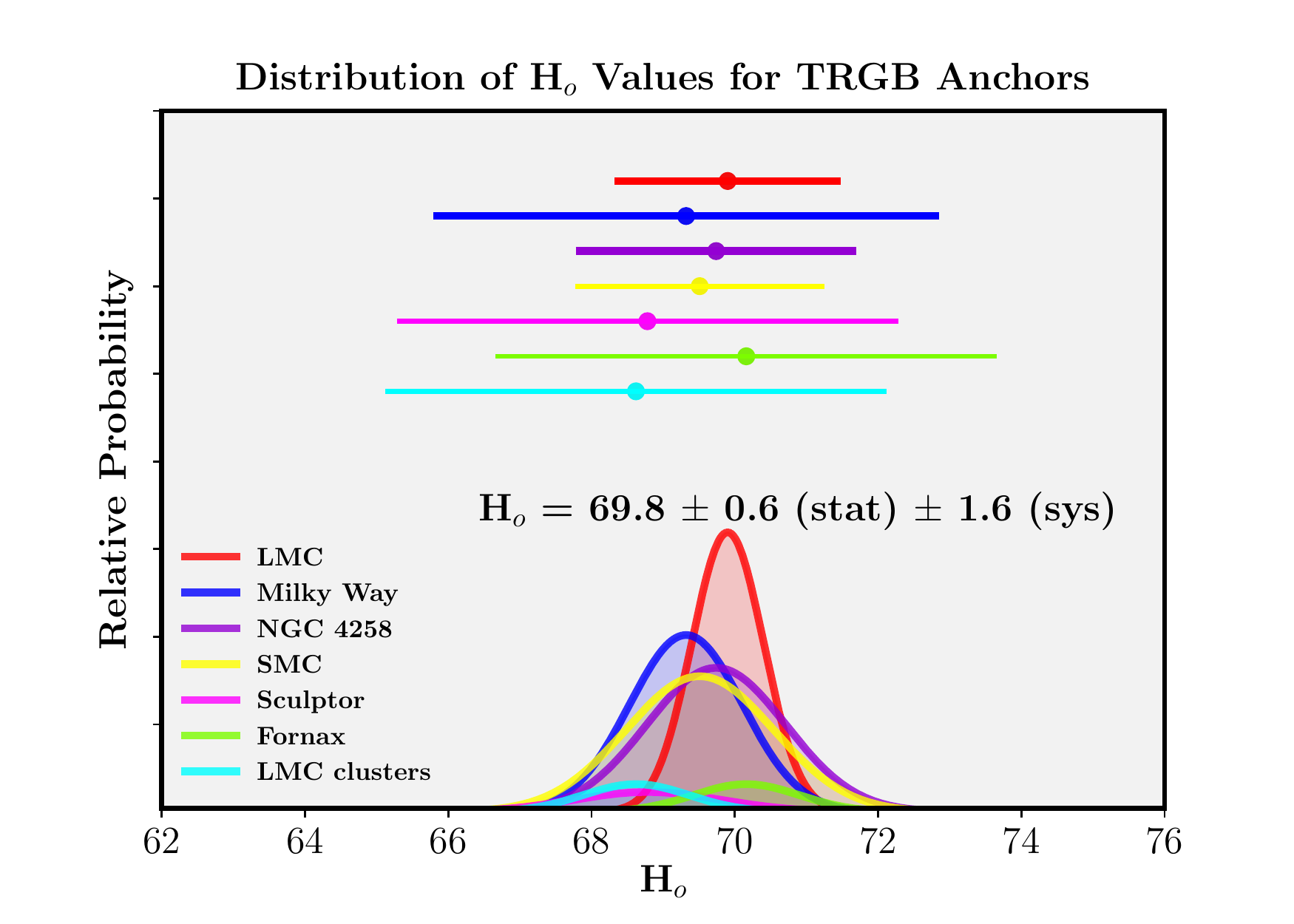} \caption{Probability density functions for the values of \ho based on the seven calibrations described in \S \ref{sec:trgbupdate}. 
 The  direct geometric calibrations for the LMC, the Milky Way, \ngc 4258, and SMC are independent of each other. The \ho values for Sculptor, Fornax and four LMC clusters are based on the Milky Way calibration of the horizontal branch (and are therefore not completely independent). They are consistent with the direct geometric calibrations, but they are not included in the final calibration.
 } 
\label{fig:PDFs_Hotrgb}
\end{figure*}

\autoref{fig:PDFs_Hotrgb} shows the PDFs for the values of \ho based on the seven calibrations of the TRGB discussed in \S \ref{sec:trgbupdate}.  The width of each   Gaussian is based on the statistical uncertainties alone for each individual determination.   The error bars at the top of the plot (using the same color coding) represent the corresponding systematic uncertainties in each case. 
 The  1$\sigma$ uncertainties are determined from the 16th and 84th percentiles for the Frequentist sum of the distributions, adding the statistical and systematic errors in quadrature:  $\sigma_{\bf i} = \sqrt{\sigma{_{stat,i}}^2 + \sigma{_{sys,i}}^2}$, and $\sigma_{\bf mean}$ =  $\sum{\sigma_{\bf i}} / \sqrt{(N-1)}$, where N = 4.
  The four objects with independent geometric distances  (the LMC, Milky Way, \ngc~4258 and the SMC) are represented by Gaussians with unit area. The secondary calibrations of Sculptor, Fornax, and the LMC clusters are based on the Milky Way calibration of the horizontal branch, and are therefore not completely independent. Once again, their areas have been scaled following \autoref{eq:scalearea} and are shown for illustrative purposes only. Thus Sculptor, Fornax and the LMC clusters do not contribute to the adopted overall calibration, but they do provide a consistency check on the horizontal-branch-to-TRGB distance scale.

From \autoref{fig:PDFs_Hotrgb}, it can also be seen that the range in the values of \ho for the various calibrators is small relative to the published systematic error bars. The small $\chi^2$ value may be indicating  that the systematic errors have been over-estimated, or, alternatively that statistical fluctuations have resulted in a fortuitously tight grouping of \ho values. In either case, a conservative estimate of the overall uncertainty still seems warranted; that is, we do not consider this (better than 1\% statistical) agreement to be indicating that \ho has now been measured to a level of 1\%.

\begin{figure*} 
 \centering
\includegraphics[width=1.0\textwidth]{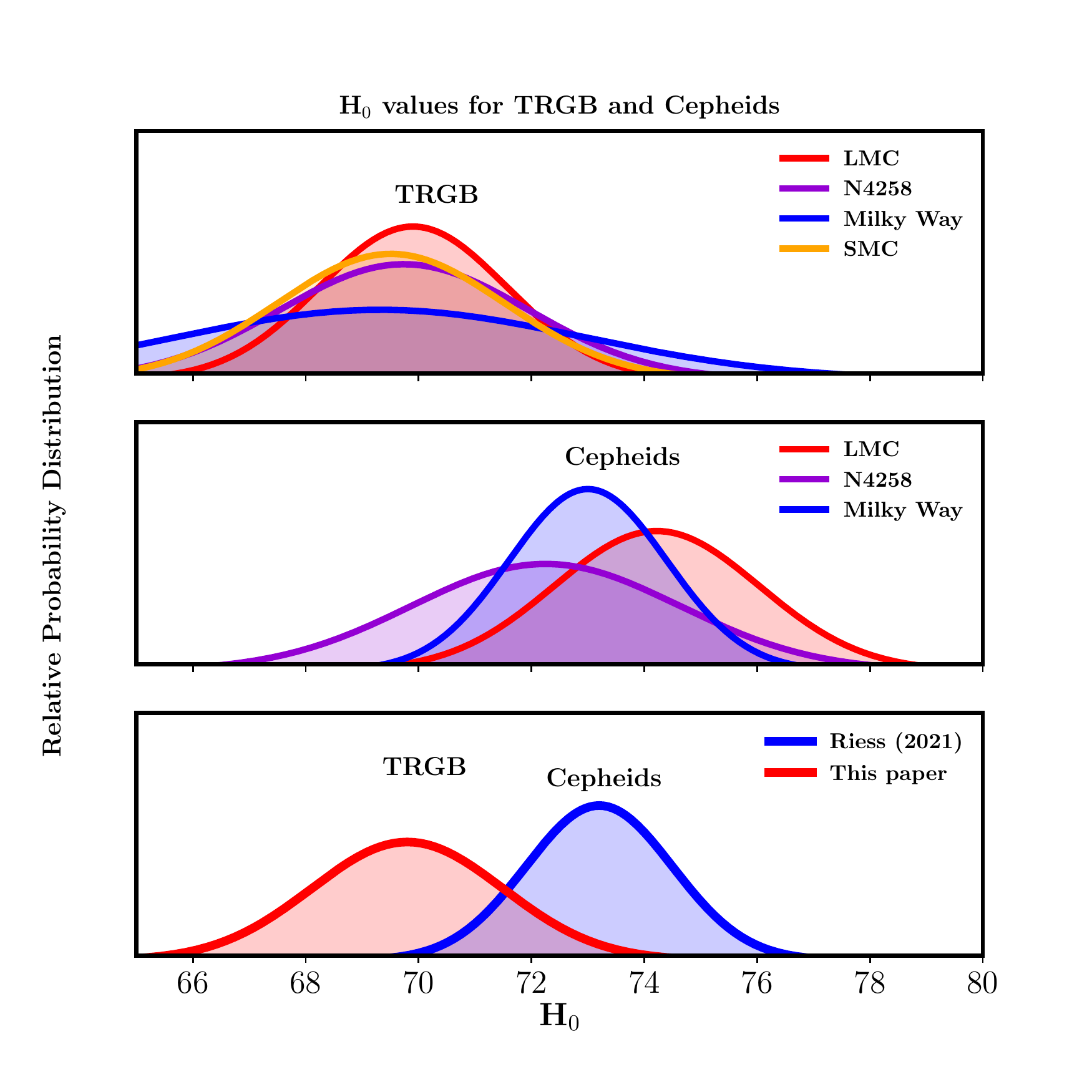}
 \caption{A comparison of the calibrations for the TRGB method and Cepheids, as listed in \autoref{tab:hotrgbceph}. Upper two panels: Probability density functions are shown for the independent calibrations for each method:  the LMC (red),  \ngc 4258 (purple), the Milky Way (blue) and the SMC (orange), in the case of the TRGB; and the LMC,  \ngc 4258, and the Milky Way in the case of Cepheids.
   Bottom panel: a comparison of the product of the probability density functions for the TRGB method and Cepheids based on the results from the upper panels.  The TRGB results are shown in red; Cepheid results are shown in blue. Note that the relative weights of the TRGB and Cepheid distributions are determined, to a large extent, by the differing uncertainties adopted for the Milky Way calibrations, where the Cepheid result assumes a highly optimistic view of the current \gaia EDR3 calibration.    }
\label{fig:PDFs_trgb_shoes}
\end{figure*}

In Figure \ref{fig:PDFs_trgb_shoes}, we show the normalized relative PDFs for the values of \ho based on the different calibrators (LMC, \ngc~4258, Milky Way, SMC), comparing both the TRGB and Cepheid calibrations in a self-consistent manner. For comparison with the \shoes results (where the statistical and systematic uncertainties are not treated independently), only the total uncertainties are considered.  The TRGB calibrations are shown at the top (in red) and the Cepheid calibrations in the middle (in blue). In this case, we follow a Bayesian approach, assuming that each anchor is equally valid, and adopting a uniform prior. The bottom panel shows the product of the PDFs. In the case of the Milky Way, the \ho values are based on the calibration from \citet{cerny_2021a} for the TRGB, and R21 for Cepheids. (The earlier Cepheid results for the Milky Way based on \hstwfciii scanning parallaxes (R16) resulted in a much higher value of \ho = 76.18 $\pm$ 2.17 \hounits.) The resulting values of \ho for the TRGB and Cepheids, respectively, are shown as solid lines. The difference between the TRGB calibration with \honewlong (this paper) and the Cepheid calibration with \ho = 73.2 $\pm$ 1.3 \hounits (R21) represents a 1.6$\sigma$  tension between the TRGB and Cepheid calibrations.

The tension between the TRGB and Cepheid calibrations is perhaps not a serious problem given that systematic uncertainties can be difficult to identify, and 2$\sigma$ is indicating generally good agreement, given those challenges. However, unlike the tension between the early universe (CMB results) and the local value of \ho, the true distances to galaxies are fixed with unique values. Rather than signifying potential new physics in the early universe, this ``local" tension is unambiguously signaling that the uncertainties in one or both distance scales (out to and including the \sne) have been underestimated.

\section{ Recent Independent Calibrations of the Cepheid Zero Point}
\label{sec:hocepheid}

\subsection{Gaia EDR3 Calibration of the Leavitt Law}
\label{sec:gaia_cepheids}

\gaia EDR3, as described in \S \ref{sec:gaia_edr3}, also presents the opportunity to derive a new zero-point calibration for Milky Way Cepheids (e.g., R21, Owens et al. 2021, in prep., Breuval et al. 2021).  (The R21
results  were shown in the third panel of Figure \ref{fig:PDFs_trgb_shoes}). We discuss below the Owens et al. \gaia EDR3-based calibration of a  multi-wavelength  sample of field Cepheids,  and compare these calibrations with the sample of field Cepheids  analyzed by R21.

Owens et al. (2021, in prep.) have analyzed \gaia EDR3 data for 49 Milky Way field Cepheids  in an attempt to provide a  multi-wavelength calibration of the Leavitt law. 
In early anticipation of the \gaia mission \citet{freedman_2011b} and  \citet{monson_2012} undertook a program to augment the sample of published optical photometry for Milky Way Cepheids with {\it Spitzer} mid-infrared (3.6 and 4.5 $\mu$m) photometry, providing a multiwavelength ($BVRIJHK[3.6][4.5]$) database for 37 Cepheids, located both in the field and in open clusters. 

Adopting the photogeometric distances obtained from the EDR3 parallax measurements by \citet{bailer-jones_2021}, Owens et al. (2021, in prep.) 
derived  optical-to-mid-infrared Leavitt law relations for the Milky Way sample. The \citeauthor{bailer-jones_2021} measurements include correction for the zero-point offset in \gaia EDR3 parallaxes \citep{gaia_lindegren_2021a}.  A challenge at present is that this sample of Milky Way Cepheids is very bright in apparent magnitude (4 $<$ G $<$ 11 mag). As already discussed in \S \ref{sec:gaia_edr3},  the corrected \gaia EDR3 parallaxes have large uncertainties, and have been shown to be underestimates. Moreover, they are significantly underestimated at brighter magnitudes \citep[e.g.,][]{el-badry_2021}, up to 30\% for isolated sources with small quoted astrometric uncertainties (and up to 80\% for those with companions). R21 found that  a --14 $\mu$as correction to their Cepheid parallaxes was indicated, obtained by minimizing the scatter in their Wesenheit Leavitt law. 

In a  comparison with \hst parallaxes and published infrared Baade-Wesselink distances, as well as the DEB distances to the LMC and SMC, Owens et al. (2021, in prep.) concluded that the current uncertainty in their sample of EDR3 parallaxes is conservatively at a level of $\sim\pm$5\%, much larger than the 1\% or better accuracy  anticipated from future (DR4 and DR5) \gaia releases. Owens et al.  also explored adding a constant offset to the Leavitt law, but found that there is no single offset that minimizes the scatter (as would be expected for distance errors) for their multiwavelength sample.  They instead  used the DEB distances measured for the LMC and SMC by \citet{pietrzynski_2019} and \citet{graczyk_2020} to provide an external estimate of the offset in the Milky Way sample, finding a value of $+$17.5 $\mu$as, similar in magnitude, but opposite in sign to that found by R21. (The sense of the offset found by Owens et al. is in the same sense as that found by \citet{maiz_apellaniz_2021}.) However, as Owens et al. emphasize, the adoption of the DEB distances does not then provide an independent \gaia EDR3 zero-point calibration, and uncertainty in the required correction to the \gaia EDR3 parallaxes remains.

Although the uncertainties are not yet at a level of 1\%, there is still internal consistency at a few percent level in the Cepheid zero points obtained using different Cepheid samples, different parallax measurements, different external constraints and analyzed by different authors. At this level, it provides evidence for stability in the Cepheid zero point, much as we saw for the internal consistency and stability in the TRGB zero point in \S \ref{sec:trgbupdate}. Once again, these results indicate that the divergence of the TRGB and Cepheid distance scales, and the resulting values of \ho, occur (at least primarily) farther out in the rungs of the distance ladder, and are not  coming from errors in the respective locally determined zero-point calibrations. 

\section{Comparison of the TRGB and Cepheid Calibrations of H$_o$}
\label{sec:comparison}

The adopted TRGB value of \honewlong  is smaller than the most recent  SHoES  Cepheid calibration at a level of $\sim$2$\sigma$. Next, we examine the implications of forcing a higher value of \ho onto the calibration of the TRGB for globular clusters in the Milky Way. 

\begin{figure*} 
 \centering
\includegraphics[width=0.7\textwidth]{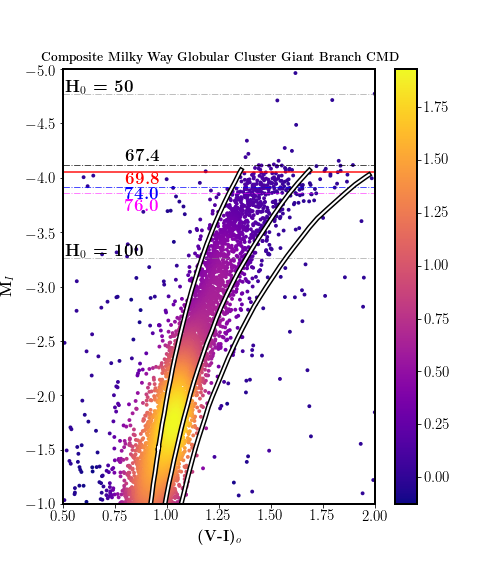}
 \caption{Composite M$_I$ versus $(V-I)_o$ color-magnitude diagram for  giant branch stars, based on a sample of 46 Galactic globular clusters, color coded by the density of points. This plot is an expansion of the red rectangle shown in \autoref{fig:GC_ALL}. The TRGB  
 is shown by the
 red line, located at an absolute $I$-band magnitude of M$_I$ = \trgbmagnew \ mag.
 This calibration results in a value of H$_o$ = \honew. 
 Shown also for comparison as blue dashed lines are the corresponding values for \ho = 67.4,  74 and 76, respectively. 
  PARSEC(Padova and Trieste Stellar Evolutionary Code: http://stev.oapd.inaf.it/cgi-bin/cmd) isochrones (CMD Version 3.3; \citet{bressan_2012}; \citet{marigo_2017})  with [Fe/H] values from left to right of -2.0, -1.2 and -0.8 dex, respectively, are illustrated by the three white curves outlined in black. The fits to these isochrones illustrates, both empirically and theoretically, how small the effect of metallicity is for the TRGB in the $I$ band at these low metallicities. The historical  \ho values of 100 and 50 are also labeled: their large spread relative to current measurements illustrates the dramatic progress in the measurement of \ho in recent decades. }
\label{fig:GC_TRGB}
\end{figure*}

Figure \ref{fig:GC_TRGB} shows an expanded version of the M$_I$ versus (V-I)$_o$ color-magnitude diagram for the Milky Way globular clusters discussed in \S \ref{sec:GC}, this time centered on the giant branch. Corresponding values of \ho are indicated. It can be seen that a value of \ho = 74 \hounits (R19) is significantly discrepant with the measured position of the TRGB, as are the values of \ho = 75-76 \hounits calibrated using scanning parallaxes for Milky Way Cepheids from R19, a recent calibration of the Tully-Fisher relation \citet{kourkchi_2020}, and surface-brightness fluctuations \citep{verde_treu_riess_2019}. Values of \ho of 74 and 76 \hounits correspond to adopting absolute magnitudes for the TRGB of M$_I$ = $-3.92$ and $-3.86$ mag, respectively, significantly fainter than virtually all calibrations  found in the published literature (see \autoref{tab:trgbzp}), and differing by 6\% and 9\% from the calibration adopted here.  Future work is required to ascertain the reason for this discrepancy,  most importantly, 1) further comparisons of individual TRGB and Cepheid distances to \sn host galaxies, and 2) the ultimate establishment of the zero point of both the TRGB and the Cepheid distance scales at a $<$1\% level with future \gaia releases. For comparison, the Planck value of \ho would correspond to adopting values of M$_I$ = $-4.12$ mag (a 3\% difference from our adopted calibration).

In \S \ref{sec:trgbupdate},  \S \ref{sec:hotrgb}, and  \S \ref{sec:hocepheid}, we have seen that recent updates to the absolute zero-points of the TRGB and Cepheid distance scales are each internally consistent with previously published zero points for each method
at the 1--2\% level, 
and therefore that the difference in the values of \ho based on these two methods cannot be (completely) ascribed to a zero-point error. A difference in \ho of 4 \hounits (i.e., between 70 and 74 \hounits) corresponds to a difference of 0.12 mag or 6\% in distance, which is about 3-6 times the quoted uncertainty in the current estimates of the TRGB and Cepheid zero points (of 1 to 2\%).

As discussed in F19, the TRGB and Cepheid distances to galaxies agree well (having a scatter of $\pm$0.05 mag or 2\% in distance) for nearby distances ($<$ 7 Mpc), but they begin to diverge for the more distant galaxies (where the scatter is over three times larger, $\pm$0.17 mag or 8\% in distance), with a  weighted average difference in distance modulus (in the sense of TRGB minus Cepheid; i.e., the TRGB distances are larger) amounting to $+$0.059 mag. Although in principle, one could adopt a TRGB zero point that is
significantly fainter than --4.05, that simply shifts the offset to (and worsens the good agreement at) closer distances where the current Cepheid and TRGB distances agree extremely well, with an average difference of $+$0.02 mag or 1\% in distance.

The strengthening of both the TRGB and Cepheid zero-point calibrations, in addition to the good agreement between the TRGB and Cepheids for distances closer than 7 Mpc, again suggests that the discrepancy in \ho arises farther afield. One potential clue as to part of the problem may be indicated by the observed scatter in the calibrated absolute \sne magnitudes, as discussed by F19. These authors found that the scatter in the TRGB-calibrated \sne magnitudes for nearby galaxies amounted to $\sigma$ = 0.11 mag, in good agreement with the scatter in the CSP Hubble diagram of  $\sigma$ = 0.10 mag for the more distant \sne sample, whereas the scatter in the Cepheid-calibrated \sne magnitudes is larger, with  $\sigma$ = 0.15 mag. Further improvement to the distances of galaxies in the 15-30 Mpc range will be needed to resolve this issue. Scheduled \jwst observations will be critical to this effort (e.g., \jwst Cycle 1 GO proposal(Proposal 01995; Freedman 2021).

In summary, the good agreement for the nearby sample suggests that {\it the zero-point calibrations of the methods are not the (primary) reason for the differences between the two methods in determining \ho}. Resolving the reason for this divergence is now critical to our understanding of whether there is new physics beyond the standard $\Lambda$CDM model.

\section{The TRGB and Cepheids as Distance Indicators}
\label{sec:discussion}

Given the historical record of large and poorly understood disagreements amongst various distance indicators (for example, the 50 versus 100 discrepancy illustrated in Figure \ref{fig:GC_TRGB}), the current (smaller) range of 67 to 74 in the value of \ho also reflects the recent significant improvement in the extragalactic distance scale. That said, in the context of testing the standard cosmological model,   it is essential to understand the origin of the difference in the TRGB and Cepheid distance scales. 

We now  turn to a discussion of each of the two methods individually. In specific, we discuss the status of the calibrations, the viability of each method as a standard candle, the effects of crowding/blending for each case, as well as the uncertainties due to dust and metallicity. We highlight the particular strengths of each method, as well as the current level of control of known systematic effects, and then outline prospects for improvement. 

\subsection{Measuring TRGB Distances}
\label{sec:trgbdist}

\begin{enumerate}

   \item {\it Calibration of the TRGB zero point}:
As shown in \S \ref{sec:trgbupdate} of this paper, direct geometric calibrations of the TRGB method for the LMC (F19, F20;  \citet{hoyt_2021b}),
\ngc 4258 \citep{jang_2021},  globular clusters in the Milky Way
\citep{freedman_2020, cerny_2021a}), and the SMC \citep{hoyt_2021b}
all agree to within $\pm$1\%.

   \item {\it The TRGB as a Standard Candle}:
The strikingly sharp and flat definition of the TRGB  at $F814W$ (comparable to the ground-based $I$ band)  for Milky Way globular clusters (see \autoref{fig:GC_TRGB}) provides growing direct evidence that old, blue metal-poor giant branch stars at the tip of the RGB are  {\it actual} standard candles, distinctive from other commonly employed {\it standardizable} candles (for example,  \sne and Cepheids). The fact that this sharp cutoff is not simply an empirical feature, but that it is the result of a well-understood physical mechanism (the core helium flash) lends confidence to the use of these stars as reliable distance indicators. 

   \item {\it Photometric Errors Due to Crowding/Blending Effects}:
The TRGB method is best applied in the outer halos of galaxies \citep[e.g., see the discussion in][and references therein]{jang_2021}, where the surface brightness of the galaxy is low, and the overlapping of stellar point spread functions is minimal. Crowding/blending effects are not currently a significant source of uncertainty for the TRGB method if carefully applied to stars in the outer halos of galaxies.

   \item {\it Effects of Dust: Foreground and Internal}: Foreground Milky Way reddening corrections are obtained from the all-sky extinction maps of \citet{schlafly_finkbeiner_2011}. Beyond the Milky Way, for the  application of the TRGB method targeted in the halos of galaxies, the effects of dust are small \citep[e.g.,][]{menard_2010}. For the four current anchors (LMC, Milky Way, \ngc~4258, and the SMC), the local line-of-sight circumstances are different for each case, and extinction and reddening corrections have been investigated in detail on a case-by-case basis as described in \citet{Jang_2018, cerny_2021a,freedman_2020} and  \citet{hoyt_2021b}. For an individual anchor, the distance uncertainty attributed to this correction contributes to its systematic uncertainty; however, for the  determination of \ho based on several anchors, it contributes only to the overall statistical error, and  not to the final systematic uncertainty.

   \item {\it Metallicity Effects}:
For red giant branch stars, there is a metallicity (and concomitant color) dependence of the luminosity that  is both predicted by theory and independently confirmed by observation  \citep[e.g.,][]{freedman_2020}. A significant advantage of the TRGB method is that it has long been known that the color of a star on the red giant branch {\it is a direct indicator of the metallicity of the star} \citep[e.g.,][]{da_costa_armandroff_1990, carretta_bragaglia_1998}. 

Given a known (flat)  TRGB slope in the $I$ band, the corresponding slope of the  giant branch luminosity with color, at any other given wavelength, is not arbitrary: it is {\it a priori} mathematically defined for the other wavelengths \citep{madore_freedman_2020}.
Empirically, the slope and zero points of the $VIJHK$ red giant branch terminations determined for the LMC and SMC agree with those measured for Milky Way globular clusters to within their 1-$\sigma$ uncertainties (F20; Cerny et al. 2020). 

For the purposes of the $I$-band ($F814W$) calibration presented in this paper, the effects of metallicity are negligible, given that only the bluest (metal-poor) stars enter the calibration, and that the flat (color-independent) nature of the TRGB {\it in this restricted color regime} is well-established (see, for example,  \autoref{fig:GC_TRGB} above, and Figure 6 of Jang et al. 2020).

\item {\it  Future Prospects for the TRGB Distance Scale}:

\noindent   
a) {\it Strengthening the Zero-Point Calibration}: 
In the future \gaia DR4  will provide twice as many observations compared to EDR3, and a new full-scale astrometric solution with a decrease in both the random and systematic uncertainties \citep{gaia_lindegren_2021a} compared to those discussed in \S \ref{sec:gaia_edr3} for EDR3.

\noindent 
b) {\it Increasing the Number of \sne Calibrators}: 

(i) As new \sne are detected in galaxies at distances  $\leq$ 30 Mpc,  \hst observations of the halos of the host galaxies can provide $I$-band TRGB distances with precisions of better than 2\% for a modest investment in telescope time. Unique in this regard is that the method can be applied to galaxies of all types, including edge-on spiral, S0 and elliptical galaxies, thus both increasing the number of calibrators and also mitigating potential systematics in the \sn data.

(ii) A combination of ground- and space-based observations can  further strengthen the  calibration of the TRGB at other (near- or mid-)infrared wavelengths \citep[e.g.,][]{dalcanton_2012, madore_2018, hoyt_2018, durbin_2020}.  Red giant stars are brighter in the near-infrared than at optical wavelengths. With \jwst, the mid-infrared TRGB calibration can be applied to distances of $\gtrapprox$40 Mpc (or a volume five times greater than currently possible with \hst), thereby adding significantly more \sne into the calibration.\footnote{This new \jwst capability is highly desirable because \sne are sufficiently rare that host galaxies for which TRGB stars (or Cepheids) are also accessible with \hst are discovered only every 1.5 to 2 years.}

\end{enumerate}

\subsection{Measuring Cepheid Distances}
\label{sec:cephdist}

\begin{enumerate}

   \item {\it Calibration of the Cepheid Zero Point}:  Direct geometric calibrations of the Cepheid Leavitt Law for the LMC are based on (a)  the DEB distance to the LMC \citep{pietrzynski_2019}; (b)  \hst and \gaia parallaxes for field Cepheids in the Milky Way \citep{benedict_2007, riess_2018,
   riess_2021}; and (c) the maser distance to \ngc 4258 \citep{reid_2019}. The  resulting  values of \ho for these three calibration methods  currently span a range of 72-74 \hounits.\footnote{\citet{efstathiou_2020} has discussed at some length the internal tension between the LMC and \ngc 4258 anchor distances (which depend upon the adopted metallicity correction), and notes that the \ho tension may be arising, in part, due to inconsistencies in the local anchors.}

   \item {\it Cepheids as Standardizable Candles}: The well-defined relationship between period, luminosity and color can, in principle, produce a standardizable candle of high precision. Then, including a metallicity term, the Leavitt law can be expressed as
   \begin{equation}
\label{eq:Leavittlaw}
M_{\lambda_1} = \alpha \ log P + \beta(m_{\lambda_1}-m_{\lambda_2})_o + \gamma [O/H] + \delta    
\end{equation}
\noindent
where the Cepheid magnitude at a given wavelength $\lambda_1$ is a function of the logarithm of the period (P), a color term with coefficient $\beta$, and a term with coefficient, $\gamma$, that allows for a metallicity effect (where [O/H] represents the logarithmic oxygen to hydrogen ratio for HII regions in the vicinity of the Cepheids, relative to the solar value); and $\delta$ is the zero point.
   
   It has long been recognized that the decreasing scatter in the correlation between period and luminosity with increasing wavelength \citep[e.g.,][]{madore_freedman_1991}, as well as the decreasing effect of reddening and metallicity with increasing wavelength, motivates the application of the Leavitt Law at near-infrared (or longer) wavelengths \citep{mcgonegal_1982, madore_freedman_1991, freedman_wilson_madore_1991, macri_2001, freedman_2008}.\footnote{An exception is the 4.5$\mu$m band in which the Cepheid flux is affected by the presence of a CO bandhead \citep{scowcroft_2011}.}

   \item {\it Photometric Errors Due to Crowding/Blending Effects}:
Cepheid variables  are yellow supergiants, generally found in relatively high-surface-brightness areas in the star-forming disks of late-type galaxies. For nearby galaxies, the crowding and blending of Cepheids is not a serious practical issue for the brightest, long-period Cepheids, but the problem worsens as the distance increases and the angular resolution decreases. Using artificial star tests, R16 concluded that these crowding/blending effects do not induce systematic effects. In addition, \citet{riess_2020}  tried to infer the quantitative effects of crowding by comparing the amplitudes of Cepheids in four galaxies out to a distance of 20 Mpc. They concluded that the erroneous measurements of Cepheid backgrounds alone cannot explain the Hubble tension. Future work is still needed to assess the implications for the even more distant galaxies in the \shoes program, which extend out to 40-50 Mpc. 

The effects of crowding/blending also become more severe with increasing wavelength where, for a given aperture telescope, the resolution is poorer in the infrared than in the optical. Disk red giants and the even-brighter asymptotic giant branch (AGB) stars (both of which are redder than Cepheids) are the main, and unavoidable, contaminants. Thus, although both dust and metallicity effects are decreasing functions of wavelength, there is a trade-off to be made with the decreasing (wavelength-dependent) resolution, and the increasing challenges of overlapping objects dominated by red stars, particularly as the distance increases.   In the case of \hst and WFC3,  the longest-wavelength available, the F160W filter (comparable to the ground-based $H$ band),  has an advantage for reducing the effects of dust and metallicity, {\it but it is at a disadvantage in dealing with the effects of increased crowding and blending}.

   \item {\it  Effects of Dust} :
As a consequence of their relative youth,  Cepheid variables are unavoidably located close to the regions of dust and gas out of which they formed. In practice, however, Cepheid reddening can be dealt with in a straightforward manner. With accurate colors, \citet{madore_1976, madore_1982}  showed that a reddening-free magnitude  can be constructed; for example, 

\begin{equation}
\label{eq:WBV}
W = V - R_V \times  (B-V), 
\end{equation}

where R$_V$ = A$_V$ /E(B-V) is the ratio of total-to-selective absorption. $W$ has been widely applied to the Cepheid distance scale \citep[e.g.,][]{freedman_2001,  riess_2016}. An advantage of $W$ is that it simultaneously corrects for all line-of-sight absorption, including both host-galaxy (internal) and Galactic (foreground) reddening.\footnote{ {As noted recently by \citep{mortsell_2021},  however, if the assumption of a universal value for R$_V$ is not valid, it could result in a systematic error in \ho, an issue that could become increasingly important in an era for which the goal is  percent level accuracy.} }

   \item {\it Metallicity Effects}:
The effects of metallicity on the Cepheid Leavitt Law are still being actively debated in the literature \citep[e.g., for a recent summary see][]{ripepi_2020}. One of the immediate challenges in constraining any metallicity effect for Cepheids is the difficulty of determining abundances for the individual Cepheids themselves.  Spectroscopic abundances have been measured  for Cepheids in the Milky Way and LMC \citep[e.g.,][]{romaniello_2008}; however, more distant Cepheids are generally too faint  to measure abundances from spectroscopy. 

Three decades of empirical tests for a Cepheid abundance effect (the measurement of $\gamma$ in \autoref{eq:Leavittlaw}  \citep[e.g.,][]{freedman_madore_1990, kennicutt_1998, romaniello_2008, fausnaugh_2015, riess_2016,  ripepi_2020, breuval_2021} have not yet led to a consensus view on the magnitude of the effect or even its sign, or indeed, whether there is an effect at any given wavelength.  Most of these studies have had to rely on the use of [O/H] abundances for nearby HII regions as a proxy for the Cepheid metallicities, which cannot generally be measured directly.  Theoretical models suggest that the effect of metallicity will be smaller at longer wavelengths, but there also remain significant differences in the predicted effects on both the slope and  intercept of the period-luminosity relations with wavelength \citep{bono_2008b,  ripepi_2020}, even at the long wavelength of the $K$ band (2.2 $\mu$m). \citeauthor{ripepi_2020} find that the slope of the metallicity term ranges from -0.04 to -0.36 mag/dex for fundamental pulsators, and from +0.23 to -0.30 mag/dex for overtone Cepheids.  Recently,  incorporating \gaia EDR3 data for the Milky Way and comparing to the LMC and SMC, \citet{breuval_2021} find that the metallicity effect is negligible in the optical ($V$ band) and moreover, contrary to previous studies, conclude that the effect {\it increases} through $IJHK$, with the largest effect being in the near-infrared.

As we enter an era where 1-2\% accuracies are required to resolve whether there is an \ho tension, it is critical that the longstanding uncertainties due to metallicity be better understood and calibrated. R16 compute a Wesenheit function of the form:

\begin{equation}
\label{eq:WVHI}
M_H^W =  m_H - R_{H,VI} \times (V-I)  {\rm ~~~~~where~~} R \equiv A_H / (A_V - A_I)
\end{equation}

\noindent
and solve for a metallicity correction on a star-by-star basis. Their conclusion is that metallicity contributes only at the  0.5\% level to their total \ho uncertainty of 2.4\%.   
Given the long-standing disagreement in the literature (both from theory and observations) further work is clearly warranted  to confirm this assertion. This issue is best addressed with multi-wavelength, high signal-to-noise data for nearby galaxies where covariant crowding effects are less severe.  

\vspace{1cm}

   \item {\it Summary and Future Prospects for the Cepheid Distance Scale}:
Cepheids have many strengths that make them good distance indicators. However, they still face a number of challenges, particularly when it comes to applying them under conditions at the limits of current telescopes and detectors, with the goal of achieving distances accurate and precise to a level of 1-2\%. The main challenge for Cepheid standardization is that several wavelengths, each of equally-high precision, are required: first to correct for reddening; second to correct for a possible metallicity effect (the wavelength dependence and sign of which remain under debate); and third, to ensure that the effects of crowding/blending are not systematically influencing the results. 

All three of the above systematic effects  (reddening, metallicity and crowding) increase toward the centers of galaxies. Since Cepheids are being crowded/blended particularly by red giant and red (even brighter) AGB stars, all three effects also will act in the sense of causing Cepheids to appear redder in regions of coincidentally higher metallicity. Put another way,  the corrections for reddening, metallicity  and crowding/blending  are covariant; for example, if the currently applied metallicity or crowding corrections are incorrect, then the reddening corrections will also be in error, because they all involve the same limited sets of colors, making it difficult to break the degeneracy. These issues will continue to pose a serious challenge for 1\% accuracy, especially when the scatter in the observed Wesenheit Leavitt law  can be 20-25\% in distance or 0.4-0.5 mag in distance modulus (R16), even for (anchor) galaxies as close as 7.6 Mpc (e.g., \ngc~4258). 
   
There are many areas where future tests could further constrain uncertainties in the Cepheid distance scale.    

\smallskip
\noindent
a) High signal-to-noise, multi-color, time-averaged ($BVIJHK$) photometry and spectroscopy for nearby galaxies with a range of metallicities can help resolve the question of the magnitude, sense and wavelength dependence of metallicity corrections.  The inclusion of additional distant galaxies will not lead to better constraints on the systematic effects such as metallicity; obtaining larger samples of galaxies will simply reduce the statistical uncertainties alone.

\smallskip
\noindent
b) As further \sne are discovered in the nearby universe, the numbers of \sne host galaxies with observable Cepheids will also slowly be increased.

\smallskip
\noindent
c) JWST/NIRCam in the $J$ band has four times the angular resolution of \hstwfciii in the $H$ band, where the longest-wavelength SHoES Cepheid measurements have been made, and thus can allow the effects of crowding/blending in the \hst photometry to be assessed directly.

\end{enumerate}

\subsection{Overall Systematics}
\label{sec:overall_systematics}

At present, the systematic accuracies of the TRGB and the Cepheid distance-scale zero-points are constrained by the small number of available geometric calibrators providing high-accuracy distances. Below we outline the degree to which the two distance scales are co-dependent (or not), on the same (or different) zero-point calibrators. 

As illustrated in  \autoref{tab:calibrators}, there are four galaxies with geometric measurements that have been used to calibrate the local distance scale: the LMC, NGC 4258, the Milky Way and the SMC. There are several important points to take away from this table. 

\begin{enumerate}

\item Both the TRGB and Cepheids adopt the same distances for the LMC and \ngc~4258, therefore sharing any systematic errors that may have been incurred in those measurements. The current total uncertainties quoted for these measurements are at a level of 1\% and 1.5\%, respectively \citep{pietrzynski_2019, reid_2019}.

\item  \ngc~4258 is the only galaxy sufficiently nearby for which an accurate maser distance can be measured, and which is also close enough for the calibration of the TRGB and Cepheids. A ``sample of one" precludes rigorous testing for potential systematic errors in this galaxy's geometric distance. 

\item  In the case of the TRGB method, the LMC and SMC calibrations share the systematic uncertainties of the \citet{skowron_2021} reddening maps. The dominant uncertainty is that of the zero point, estimated by \citet{hoyt_2021b} to be $\pm$ 0.014 mag (0.6\%) and $\pm$ 0.018 mag (0.8\%), respectively. In addition, their DEB distances are both based on the surface-brightness-color relation from \citet{pietrzynski_2019}, estimated to be 0.8\%. 

\item Finally, it should be noted that for both the TRGB and Cepheids, the same reddening law is adopted and assumed to be universal; moreover, the same ratio of total-to-selective absorption, $R_V$, is adopted in both applications. However, the TRGB method is less susceptible to the assumption of the universality of the reddening law because the dust content in the halos of galaxies is generally neglible compared to that in the disks. 

\end{enumerate}

\begin{deluxetable*}{lcc}
\tablecaption{Zero-Point Calibration  \label{tab:calibrators}} 
\tablehead{\colhead{Calibrator} & \colhead{TRGB
} & \colhead{Cepheids} } 
\startdata
LMC  & DEB\tablenotemark{a} &  DEB\tablenotemark{a}  \\
\ngc~4258  & masers\tablenotemark{b}  &   masers\tablenotemark{b} \\
Milky Way  & $\omega$ Cen DEB\tablenotemark{c}  &   EDR3 parallaxes\tablenotemark{d}  \\
SMC  &  DEB\tablenotemark{e}  &   ...  \\
\enddata
\tablenotetext{a}{~\citet{pietrzynski_2019}}
\tablenotetext{b}{~\citet{reid_2019}}
\tablenotetext{c}{~\citet{thompson_2001}}
\tablenotetext{d}{~\citet{riess_2021}}
\tablenotetext{e}{~\citet{graczyk_2020}}
\end{deluxetable*}

\subsubsection{The \sn Host Galaxies}

The tie-in to the more distant SN Ia host galaxies is similarly limited by the fact that \sne in the local universe are rare. As noted previously, there are currently 19 published Cepheid distances to \sn hosts, an equal number for which there is a TRGB calibration, and a sample of 10 galaxies for which there is an overlap. Any peculiarities in the \sne in this overlap sample (that are not shared by the more distant supernovae in the Hubble flow) will carry covariant systematics into the TRGB and Cepheid \ho determinations. 

Once again, the same reddening law is adopted for both the TRGB and Cepheids. It is assumed to be universal, and the same value or $R_V$ is adopted in both applications. However, the explicit Galactic foreground reddening corrections  used for the TRGB are decoupled from the Cepheid de-reddening process that implicitly corrects for total line-of-sight reddening using the Wesenheit method. 

\subsubsection{Distant \sne in the Hubble Flow }

Both the TRGB and Cepheids tie in to more distant \sne in the Hubble flow for the final step in the determination of \ho. While different filters, different software analysis tools, and different groups have analyzed the data, any unknown systematics in \sn distances will be shared by both methods. However, the TRGB method, which can be applied to both elliptical and spiral galaxies, will be less sensitive to correlations that are host-galaxy mass dependent.

As the distances to more \sn host galaxies are measured using \hst and \jwst, the statistical (uncorrelated) errors will decrease as 1 / $\sqrt(N)$. As the above discussion makes clear, however, there are parts of the systematic error budgets for the TRGB and Cepheid \ho determinations that are covariant.  Unfortunately, quantifying these potential covariant effects (many of which fall into the category of current unknowns: e.g., reddening laws, unknown systematics in masers, DEBs, \sne etc.) is not a realistic prospect.   Ultimately, completely independent methods (e.g., gravitational wave sirens) will be required to test for and place external constraints on covariant systematics in the  local distance scale.

\rm

\section{Comparison With Other Recent Determinations of $H_o$}
\label{sec:ho_recent}

To date, only ten \sne host galaxies have both TRGB and Cepheid distances measured (F19). Yet this sample is significantly larger than available for any other primary distance indicator. Stated another way, this is the first independent and direct test {\it for individual galaxies} in the Cepheid-supernova distance scale, and significant differences between the TRGB and Cepheid distances have been found. What about other tests? 

Although a case has been made that there are many independent checks of the Cepheid distance scale \citep[e.g.,][]{verde_treu_riess_2019},  the small number of galaxies currently available precludes a detailed and direct comparison of the  Cepheid distance scale with most other distance indicators.  For example,  the Mira method \citep{huang_2018}  is currently based upon the detection of these long-period variable stars  in  a single galaxy,  \ngc 4258, calibrated via masers in that galaxy. Furthermore, the calibration of \ho using this method then relies on observations of a single  \sn host galaxy, \ngc 1559 \citep{huang_2020}.  

Similarly, \ngc 4258, at a distance of 7.6 Mpc, is the only galaxy in the nearby universe  where the host is close enough to  have a measured Cepheid distance where the maser technique can also be applied  \citep{reid_2019}.  Furthermore, there are only six galaxies in total (including \ngc~4258) for which maser distances have been measured and used to estimate \ho, \citep{pesce_2020}; the statistical errors for this technique are thus still large compared with, for example, \sne \citep[\sne;][]{scolnic_2018, burns_2018} where samples of a hundred or hundreds of \sne have been measured. The five additional megamaser galaxies beyond \ngc~4258 have distances ranging from 50 to 130 Mpc (with recession velocities of 680 to 10200 km/s),  and peculiar-velocity corrections remain a significant source of uncertainty. (An average peculiar velocity correction of 250 km/s is about 30\% of the recession velocity of 679 km/s for \ngc~4258.)  For the total sample of six galaxies, \citeauthor{pesce_2020} find  values of \ho ranging from 71.8 to 76.9 \hounits, depending on what assumptions are made, and/or which models are adopted for the peculiar velocities.

\begin{figure*} 
 \centering
\includegraphics[width=1.0\textwidth]{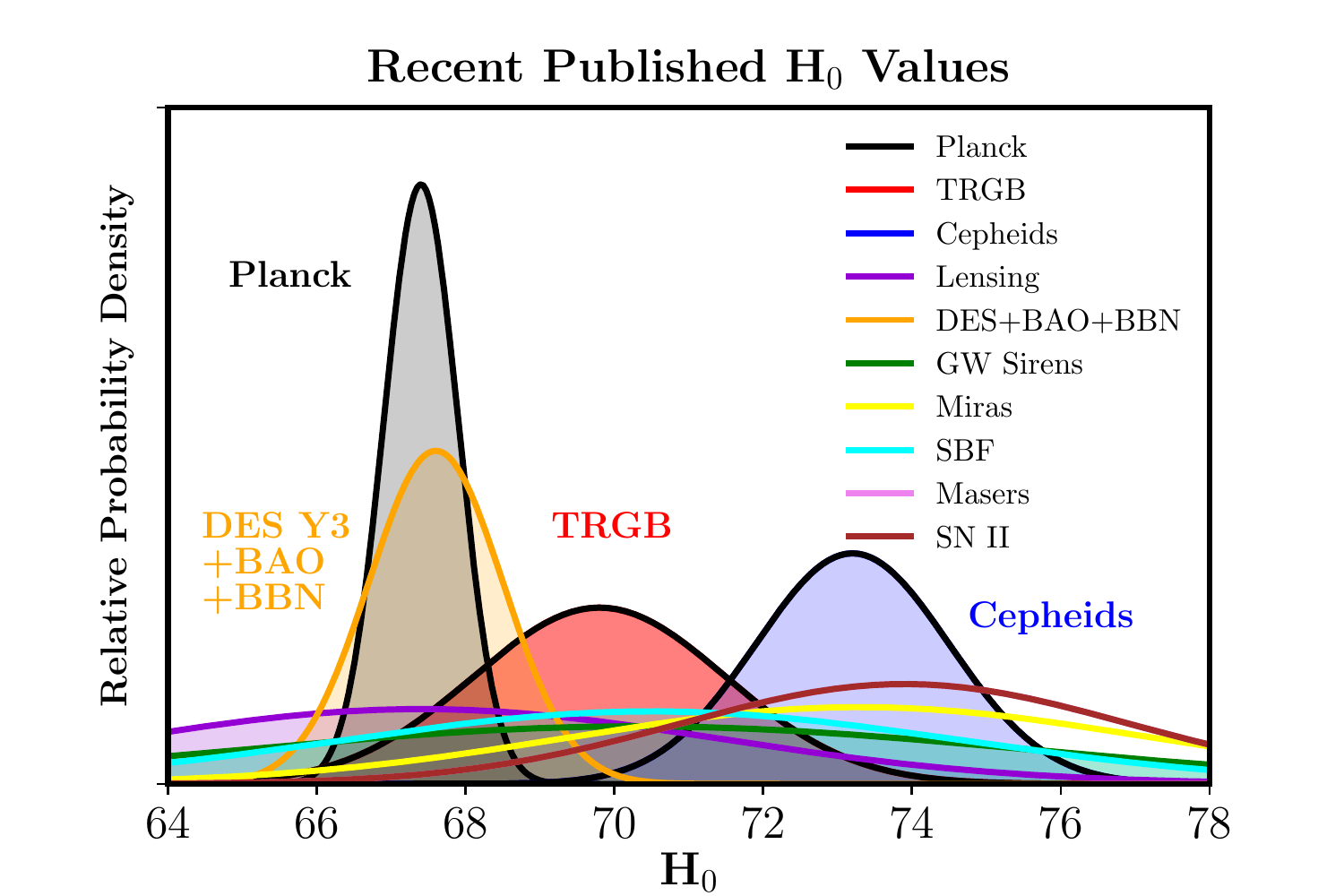}
 \caption{Relative probability density functions for several current methods for measuring \ho. The CMB, BAO, strong lensing and TRGB methods currently yield lower values of \ho, while Cepheids  yield the highest values. The uncertainties associated with \ho measurements from gravitational wave sirens, strong lensing, Miras, masers, and SBF are currently significantly larger than the errors quoted for the TRGB and Cepheids. See text for details. (CMB: Planck Collaboration 2018;   TRGB: this paper; Cepheids: R21;
 Lensing: \citet{birrer_2020};  DES Y3 + BAO + BBN: 
 \citet{DES_Abbott_2021}; GW sirens:
 \citet{hotokezaka_2019}
Miras: \citet{huang_2018}; SBF: \citet{khetan_2021}; Masers: \citet{reid_2019}).}
\label{fig:PDFs_all}
\end{figure*}

In \autoref{fig:PDFs_all}, we show a comparison of several recent  determinations of \ho and their published uncertainties. Plotted are the relative probability density functions color-coded as labeled in the legend and include:  the TRGB (this paper); Cepheids (R21);
those based on early-universe measurements (CMB: \citet{planck_2018},  the Dark Energy Survey Year 3 + BAO + BBN
\citep{DES_Abbott_2021};
as well as  gravitational wave sirens 
\citep{hotokezaka_2019}; 
Miras \citep{huang_2020}; surface brightness fluctuations  \citep[SBF][]{khetan_2021, blakeslee_2021}; masers \citep{reid_2019}; and recent results from strong lensing \citep{birrer_2020}. The Planck, DES Year 3 +  BAO + BBN, TRGB and Cepheid PDFs are also explicitly labeled.

From this figure, the discrepancy between the early universe (CMB + BAO) and local Cepheid measurements of \ho is apparent, as is the difference between the TRGB and Cepheid local determinations. Both the TRGB and Cepheid measurements have smaller uncertainties than the other (local) methods shown. These two methods currently have the largest samples of nearby objects (19 in both cases) that tie directly into the Hubble flow via \sne. 

Thus, the current situation is that there are two different types of  tensions in play: 1) that between Cepheid measurements and the early universe   and  2) that between Cepheid measurements and the TRGB. 

For completeness, in Appendix \ref{app:appendix_steer}, we show a plot of 1,065 \ho values as a function of time for  published data since 1980 (Ian Steer, private communication), as well as their histogram distribution. Interestingly, there is no bimodality (67 versus 73) seen in the overall distribution of the recently published \ho values, as can be seen in \autoref{fig:Steerhist}.

\section{Summary }
\label{sec:summary}

In this paper, we have provided an update on the calibration of the  absolute $I$-band magnitude of the TRGB anchored using several independent geometric zero points. This updated calibration includes 1) extensive measurements of the TRGB over a wide area in the halo of the maser galaxy \ngc 4258 \citep{jang_2021};  
2) independent observations of the TRGB in 46 Milky Way globular clusters covering a wide range of metallicities (Cerny et al. 2020);
and 3) a reanalysis of the TRGB incorporating revised reddening corrections for the LMC and SMC \citep{hoyt_2021b}. These calibrations all agree with that earlier determined for the LMC alone (F19, F20) to better than 1\%, providing multiple consistency checks on the LMC calibration of F19 and F20. Each of these calibrations is tied to geometrical distance anchors (H$_2$O megamasers in the case of \ngc 4258;   DEB distances and \gaia EDR3 parallaxes for the Milky Way globular clusters; and DEB distances for the LMC and the SMC). In addition, using a fiducial horizontal branch sequence defined by the Milky Way globular clusters, we discuss and compare the TRGB absolute magnitude for the  nearby dwarf elliptical galaxies Sculptor (Tran et al. 2021, in prep) and Fornax (Oakes et al. 2021, in prep), and for four LMC globular clusters,  finding  excellent additional agreement.

An improved value of \ho is determined by applying this new TRGB calibration to a sample of distant \sne. This measurement is based on: 1) the new calibration of the absolute $I$-band magnitude of the TRGB (\trgbmagnewlong) presented in this paper; 2) \hstacs observations of TRGB stars in the halos of nearby galaxies known to host \sne (F19, F20, \citet{hoyt_2021a}); 3) a sample of 99 well-observed \sne with multiwavelength photometry from the CSP \citep{krisciunas_2017}. Our final adopted value is 
\begin{equation} 
\label{eq:Hovalue}
  H_o = 69.8 \pm 0.6 \ \mathrm{(stat)} \pm 1.6 \ \mathrm{(sys)} \ \mathrm{ km s^{-1} Mpc^{-1}} .
\end{equation}
\noindent
This value of \ho, based on the TRGB, agrees to within 1.3$\sigma$ with that inferred from modeling of the CMB observations.

Currently the TRGB method and Cepheids provide the largest (statistically robust) and strongest (tested for systematics) base of distance determinations for the calibration of \ho in the local universe. 
Together they provide a check on the overall systematics. It is a testament to each method that a comparison for the nearest galaxies (i.e., within 7 Mpc) agrees in both zero point and scatter to better than 2\% accuracy (F19). However, these same two distance scales diverge  at larger distances. It is important to understand the source of this divergence, and to ascertain whether its resolution will strengthen or weaken the case for additional physics. The fact that any given galaxy must have a unique distance means that systematic errors in one or both of the current estimates must be the cause for the divergence. At this time, the outcome is unknown: no clear evidence for outstanding systematic effects in either the TRGB or Cepheid distances has been found.  It should be noted, however,  that crowding/blending effects are not an issue for the TRGB, that multiple geometric determinations of the zero point show consistency at the 1\% level, and that metallicity effects are  better understood from theory, and more easily addressed empirically, for TRGB stars than for Cepheids. Finally,  a number of ongoing studies of the TRGB and Cepheids, combined with the upcoming launch of \jwst, plus improvements to the \gaia zero points in future releases, all hold promise for significant improvement leading to a  resolution of the current discrepancies within the next few years.

\begin{figure*} 
 \centering
\includegraphics[width=1.0\textwidth]{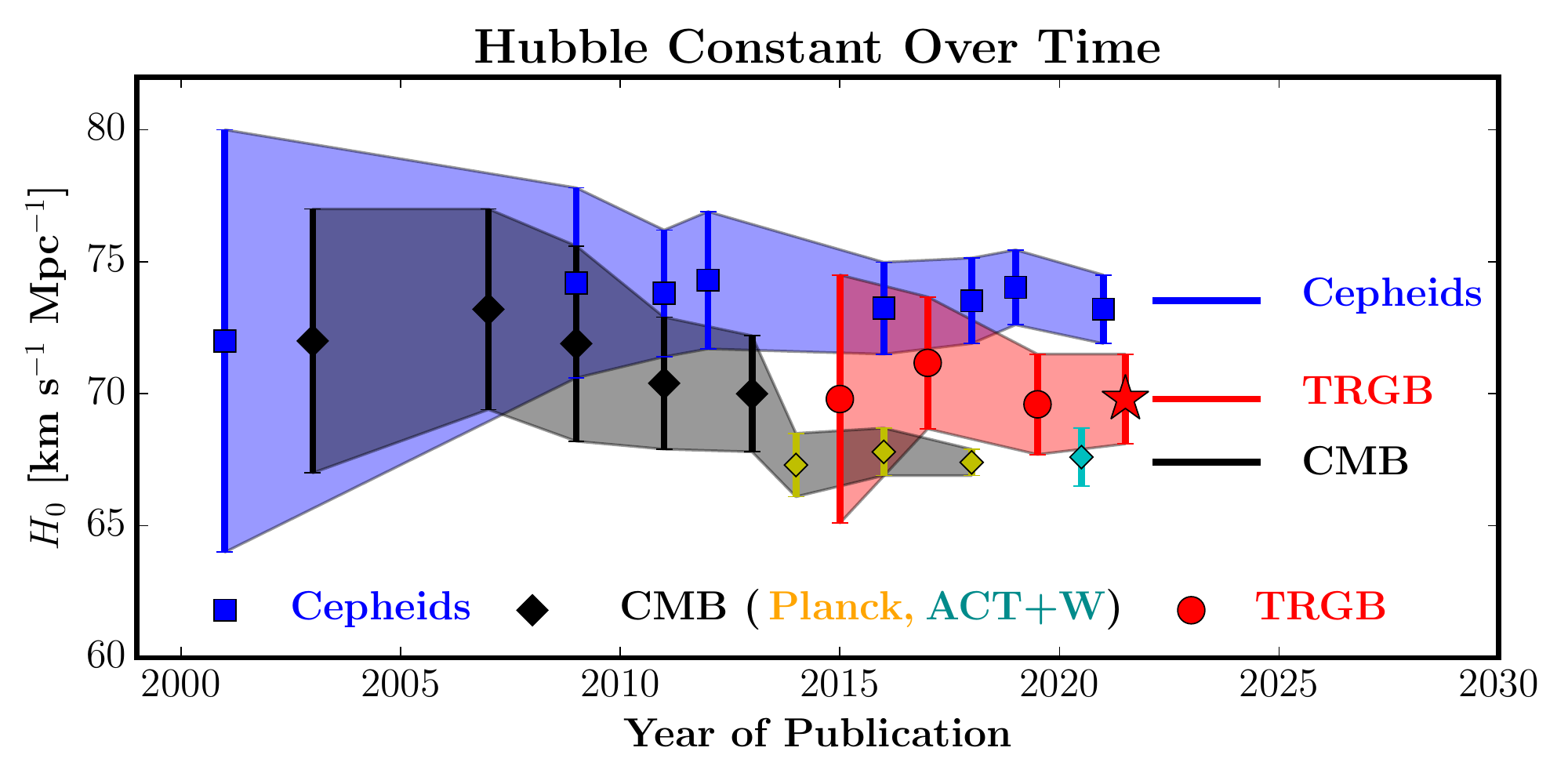}
 \caption{A summary of Hubble constant values in the past two decades, based on Cepheid variables (blue squares), the TRGB (red filled circles and star), and estimates based on measurements of fluctuations in the CMB (WMAP: black filled diamonds; Planck: yellow diamonds; ACT + WMAP: cyan diamond). 
 The  CMB \ho values assume a flat $\Lambda$ CDM model. 
 The CMB and Cepheid results straddle a range of 67 to 74 \hounits, with the TRGB results falling in the middle, and overlapping the CMB results. The tension between the CMB and TRGB results amounts to only  1.3$\sigma$. }
\label{fig:hotime}
\end{figure*}

\acknowledgements

Support for program \#13691  was provided by NASA through a grant from the Space Telescope Science Institute, which is operated by the Association of Universities for Research in Astronomy, Inc., under NASA contract NASA 5-26555.
The \cspi has been supported by the National Science Foundation under grants AST0306969, AST0607438, AST1008343, AST1613426, and AST1613472.
Computing resources used for this work were made possible by a grant from the Ahmanson Foundation.
This research has made use of the NASA/IPAC Extragalactic Database (NED), which is operated by the Jet Propulsion Laboratory, California Institute of Technology, under contract with the National Aeronautics and Space Administration.
Some of the data presented in this paper were obtained from the Mikulski Archive for Space Telescopes (MAST). STScI is operated by the Association of Universities for Research in Astronomy, Inc., under NASA contract NAS5-26555. 
I thank the {\it Observatories of the Carnegie Institution for
Science} and the {\it University of Chicago} for their support of long-term research into the calibration and determination of the expansion rate of the Universe. 
My thanks to many collaborators who have contributed to various facets of this research on the TRGB, Cepheids and supernovae; in particular Barry Madore for his many decades of collaboration on the distance scale;    as well as current and previous students  Taylor Hoyt,  William Cerny, Quang Tran, Elias Oakes, Kayla Owens, Finian Ashmead,  and Dylan Hatt;  postdoctoral fellows In Sung Jang,  Rachael Beaton and Jill Neeley; research scientists and faculty Andy Monson, Mark Phillips, Mark Seibert, Jeff Rich and Myung Gyoon Lee. Special thanks to Chris Burns for re-running his SNooPy and STAN MCMC code on the CCHP TRGB sample updated since 2020, and for creating \autoref{fig:SNehists}. In addition, I gratefully acknowledge Ian Steer for providing access to his \ho database. 
I thank Barry Madore, Kayla Owens, In Sung Jang and Taylor Hoyt for their comments on the manuscript, as well as an anonymous referee for several helpful suggestions to update and improve the paper.

{\it Facilities:} $HST$ ($ACS$) $Gaia$

\software{
          Matplotlib, 
          NumPy, 
          SciPy 
          }

\bibliographystyle{apj}
\bibliography{ms}

\appendix 

\section{Hubble Constants Published Since 1980}
\label{app:appendix_steer}

\begin{figure*}
\figurenum{A1}
\centering
\includegraphics[angle=0,width=1.0\textwidth]{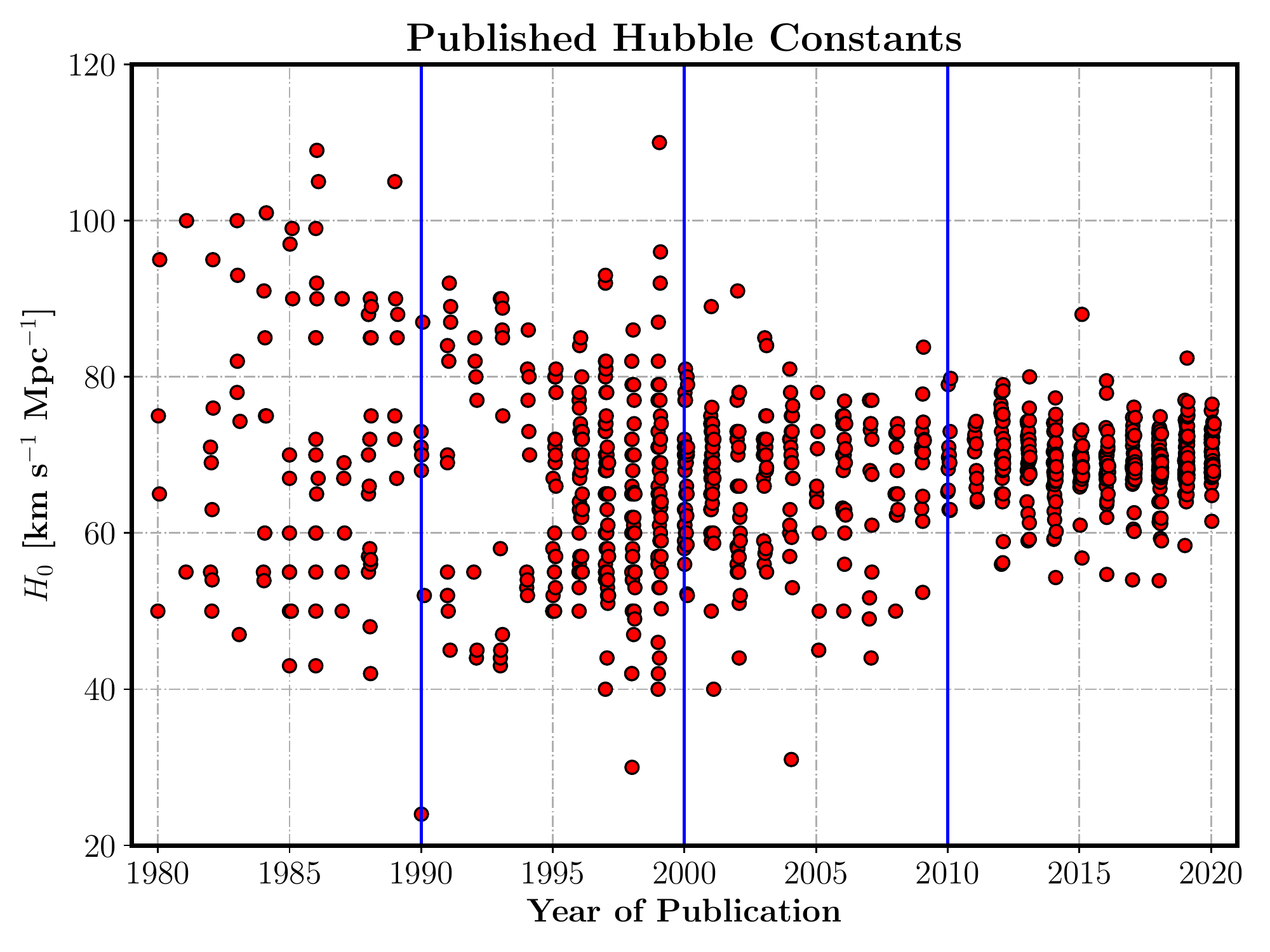}
\caption{Plot of published \ho values since 1980. 
Data courtesy of Ian Steer, private communication. These data provide an update of the John Huchra Hubble constant database originally maintained for the NASA Hubble Space Telescope Key Project on the extragalactic distance scale \citep{freedman_2001}. This figure further updates that shown in \citet{steer_2020} with an additional 99 entries.
\label{fig:Steer}}
\end{figure*}

\autoref{fig:Steer} plots \ho values   published since 1980. The scatter in published \ho values has continued to decrease with time.  All methods are included, without judgement  as to accuracy of a given method. In this sense it is an unbiased sample. 

\begin{figure*}
\figurenum{A2}
\centering
\includegraphics[angle=0,width=1.0\textwidth]{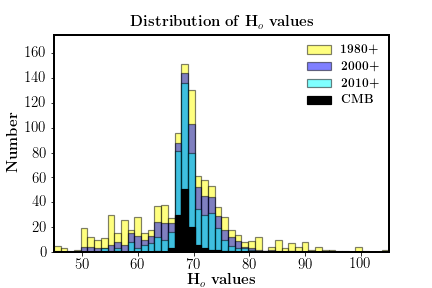}
\caption{Histogram distributions of \ho values for all published data since 1980 (yellow), data since 2000 (purple), data since 2010 (cyan) and Ho estimates from CMB data (black). Data source is the same as for \autoref{fig:Steer}.
\label{fig:Steerhist}}
\end{figure*}

\autoref{fig:Steerhist} shows in histogram form the distribution of \ho values. It illustrates clearly how the scatter in \ho values has decreased over the past four decades. For the most recent decade (2010-2020), the average, median and mode of the \ho distribution are 68.9, 68.6 and 68.0 \hounits, respectively. The values of \ho inferred from measurements of the CMB are shown in black.  Interestingly, no obvious bimodality of \ho values is seen between the values of 67 and 74 \hounits, the two values that define the current ``\ho tension". 

\end{document}